\documentclass[10pt]{article}
\usepackage{fullpage,graphicx,subfigure,mathdots,mathpazo,color}
\usepackage{amsmath,amscd,tikz,mathrsfs,cite}
\usepackage[normalem]{ulem}
\usepackage{amsmath}
\usepackage{setspace}

\usepackage{epsfig,amsmath,graphicx,amssymb,overpic}
\usepackage{bm}
\usepackage{graphicx}
\usepackage{subfigure, xcolor}
\usepackage{ntheorem}
\usepackage{listings,diagbox}

\usepackage{color}
\usepackage{epsfig,amsmath,graphicx,amssymb,overpic}

\usepackage{booktabs}
\usepackage{diagbox}
\usepackage{graphicx}
\usepackage{dcolumn}
\usepackage{bm}
\usepackage{graphicx}
\usepackage{subfigure}
\usepackage{epsfig,amsmath,graphicx,amssymb,overpic,cite}
\usepackage{makecell}

\usepackage{epsfig,amsmath,graphicx,amssymb,overpic}
\usepackage{bm}
\usepackage{graphicx}
\usepackage{subfigure, xcolor}
\usepackage{float}
\usepackage{listings}
\usepackage{tabularx}
\usepackage{multirow}

\def\be{\begin{equation}}
\def\ee{\end{equation}}
\def\bee{\begin{eqnarray}}
\def\ene{\end{eqnarray}}
\def\bes{\begin{subequations}}
\def\ees{\end{subequations}}

\newcommand{\bx}{{\bm x}}

\newcommand{\PT}{{\cal PT}}
\newcommand{\T}{{\cal T}}

\def\v{\vspace{0.1in}}

\def\be{\begin{equation}}
\def\ee{\end{equation}}
\def\bee{\begin{eqnarray}}
\def\ene{\end{eqnarray}}
\def\bes{\begin{subequations}}
\def\ees{\end{subequations}}

\def\d{\displaystyle}

\def\v{\vspace{0.1in}}

\usepackage{geometry}
\allowdisplaybreaks[4]

\begin{document}

\baselineskip=14pt
\renewcommand {\thefootnote}{\dag}
\renewcommand {\thefootnote}{\ddag}
\renewcommand {\thefootnote}{ }

\pagestyle{plain}


\begin{center}
\baselineskip=16pt \leftline{} \vspace{-.3in} {\Large \bf Deep learning soliton dynamics and complex potentials recognition for 1D and 2D $\PT$-symmetric saturable nonlinear Schr\"{o}dinger equations} \\[0.2in]
\end{center}


\begin{center}
Jin Song$^{1,2}$ and Zhenya Yan$^{\rm 1,2,*}$
\footnote{$^{*}${\it Email address}: zyyan@mmrc.iss.ac.cn (Corresponding author)}
\\[0.05in]
{\small $^1$KLMM, Academy of Mathematics and Systems Science, Chinese Academy of Sciences, Beijing 100190, China \\
$^2$School of Mathematical Sciences, University of Chinese Academy of Sciences, Beijing 100049, China} \\
\end{center}



\noindent {\small {\bf Abstract.}
In this paper, we firstly extend the physics-informed neural networks (PINNs) to learn data-driven stationary and
non-stationary solitons of 1D and 2D saturable nonlinear Schr\"{o}dinger equations (SNLSEs) with two fundamental $\PT$-symmetric Scarf-II and periodic potentials in optical fibers. Secondly, the data-driven inverse problems are studied for  $\PT$-symmetric potential functions discovery rather than just potential parameters in the 1D and 2D SNLSEs. Particularly, we propose a modified PINNs (mPINNs) scheme to identify directly the $\PT$ potential functions of the 1D and 2D SNLSEs by the solution data.
And the inverse problems about 1D and 2D $\PT$-symmetric potentials depending on propagation distance $z$ are also investigated using mPINNs method.
We also identify the potential functions by the PINNs applied to the stationary equation of the SNLSE.
Furthermore, two network structures are compared under different parameter conditions such that the predicted $\PT$ potentials can achieve the similar high accuracy. These results illustrate that the established deep neural networks can be successfully used in 1D and 2D SNLSEs with high accuracies. Moreover, some main factors affecting neural networks performance are discussed
in 1D and 2D $\PT$ Scarf-II and periodic potentials, including activation functions, structures of the networks, and sizes of the training data.
In particular, twelve different nonlinear activation functions are in detail analyzed containing the periodic and non-periodic functions such that it is concluded that selecting activation functions according to the form of solution and equation usually can achieve better effect.}

\vspace{0.1in} \noindent  {\bf\it Keywords:} \, Deep neural network learning,\, \, Soliton dynamics, Complex potentials recognition,\, 1D and 2D saturable nonlinear Schr\"{o}dinger equation, \,$\PT$-symmetric non-periodic and periodic potentials





\baselineskip=13pt

\section{Introduction}\label{sec1}

\quad As is well-known, soliton dynamics plays an important role in many fields of nonlinear sciences, such as nonlinear optics,
Bose–Einstein condensates, plasma physics, fluid mechanics, and even finance~\cite{soliton1,soliton2,soliton3,mal19,rc1,rc2}. Since Bender,
{\it et al}\cite{pt} proposed the concept of ${\mathcal PT}$ symmetry in 1998, the $\PT$-symmetric solitons have drawn more and more attention in the field of nonlinear sciences, such as nonlinear optics, quantum optics, Bose-Einstein condensates, material science, etc. (see Ref.~\cite{pt8} and reference therein).
 Non-Hermitian Hamiltonians including $\PT$-symmetric potentials admit fully real (physically meaningful) spectra. And the symmetry is realized in this case with parity operator $\mathcal{P}$ and time-reversal one $\T$ defined as $\mathcal{P}: \bx\rightarrow -\bx$; $\T: i\rightarrow -i,\,z\rightarrow -z$.
Until now, plenty of $\PT$-symmetric stable solitons and many new nonlinear wave phenomena have been discovered in the nonlinear physical models, such as the nonlinear Schr\"{o}dinger equation (NLSE) with $\PT$-symmetric potentials~\cite{pt1,pt2,pt3,pt4,pt5,pt6,pt7,pt10,pt8,We15,ptc1,ptc2}.
 The known $\PT$-symmetric potentials include the Scarf-II potential, harmonic potential, Gaussian potential, Rosen-Morse potential, optical lattice potential, and others \cite{pt7,ptc1,ptc2,ptsc1,ptha1,ptga1,ptrm1}. 
Especially, saturable nonlinearity (SN) has advantages over other general nonlinear terms such as the Kerr one in semiconductor doped glasses~\cite{semicon} and photorefractive media~\cite{semicon2}. In two and three dimensions the collapses of fundamental solitons can be suppressed by SN \cite{sn1,sn2},
which facilitates our study of stable solitons in multidimensional optical beams. In fact, there have been many studies on the NLSE with SN and real or complex $\PT$-symmetric potentials~\cite{sn-1,sn-2,sn-3,sn-4,sn-5,sn-6,sn-7,sn-8,sn-9}.

In the field of scientific computing, traditional numerical methods such as the finite difference method and the finite element method can be replaced with a neural network to approximate the solution of partial differential equations (PDE) with the aid of automatic differentiation methods~\cite{ad,ad1}, which reduce the cost of constructing computationally-expensive grids. Recently, being different from typical data-driven deep learning (DL) methods, the physics-informed neural networks (PINNs) approach~\cite{pinn} was used to consider the important physical laws given by the PDE to control the output solution of a deep neural network. And the PINNs method has been extended to solve the stochastic differential equations, fractional differential equations, and integro-differential equations~\cite{fractional,stochastic,stochastic1,deepxde}, and soliton equations~\cite{Li20,Li21,Wang-pd22}.
Furthermore, solving inverse problems has been a hot topic.
However, solving inverse problems often costs more than solving the forward problems. Often complicated formulations, new algorithms and elaborate computer codes are required.
It is convenient for PINNs to solve inverse problems because it requires only minimum changes for the codes of forward problems \cite{pinn,inverse1}.
And recently, a Python library DeepXDE was shown for some PINNs approaches to solve multiphysics problem, which can make the
codes stay  manageable and compact~\cite{deepxde}.

 More recently, the forward and inverse problems for the cubic and  logarithmic NLS equations with $\PT$-symmetric harmonic, Scarf-II,  Gaussian, periodic, Rosen–Morse potentials potentials have been solved via the PINN method~\cite{Zhou-pla21,Wang-pla21,li21,chaos22,zhong22}.
Some improved PINN algorithms were used to study data-driven solutions for classical integrable systems \cite{chenc1,chenc2,chenc3,chenc4}.
As is well-known, the saturable nonlinearity and complex $\PT$-symmetric potentials play the important roles in nonlinear optics and other fields~\cite{sn-1,sn-2,sn-3,sn-4,sn-5,sn-6,sn-7,sn-8,sn-9}. In this paper, we would like to investigate data-driven stationary and non-stationary solutions, as well as complex potential discovery for the 1D and 2D saturable NLS equations (SNLSEs) with two fundamental $\PT$-symmetric potentials (i.e. Scarf-II and periodic potentials)~\cite{sn-5, sn-9}
\begin{equation}\label{nls}
   i\psi_z+\nabla_{\bx}^2\psi+[V(\bx)+iW(\bx)]\psi+\frac{g|\psi|^2\psi}{1+S|\psi|^2}=0,\quad g=\pm 1,
\end{equation}
by applying the deep learning PINNs and its modification, where $\psi=\psi(\bx, z)$ denotes the complex envelope field, the subscripts denote the partial derivatives with respect to the spatial variables $\bx$ and $z$ representing the propagation distance of light beam, $\nabla_{\bx}$ is the gradient operator (e.g., $\nabla_{\bx}=(\partial_x, \partial_y)$ for the 2D case), and $g$ denotes the self-focusing ($g = 1$) or defocusing ($g = -1$) nonlinearity. The non-negative parameter $S$ represents the degree of saturable nonlinearity. The complex potential $V (\bx)+iW(\bx)$ is considered to be $\PT$-symmetric provided that $V (\bx) = V (-\bx)$ and $W(-\bx) = -W(\bx)$.  Eq.~(\ref{nls}) is associated with a variational principle $i\partial\psi/(\partial z)=\delta H/(\delta\psi^*)$ with the Hamiltonian
\bee
H=\int_{\mathbb{R}}\left[|\nabla_{\bx}\psi|^2-[V(x)+iW(\bx)+g/S]|\psi|^2+\frac{g}{S^2}\ln(1+S|\psi|^2)\right]dx.
\ene

The novelties of our study are summarized as follows: on the one hand, as we know, the activation function is one of the important features of NN, which determines the activation of specific neuron during learning process. As a matter of fact, there is no clear rule on how to choose the more powerful activation function. The choice of activation function often depends on the issue itself. Therefore, the first novelty of our study is that
we introduce some new activation functions, namely ${\rm sech}(x)\tanh(x),$\, ${\rm sech}^2(x)$, $1/(1+x^2),\, x/(1+x^2)$, and test their abilities of the network performances by comparing them with other known activation functions, such as ReLU$(x)$, ELU$(x)$, Sigmoid$(x)$, Swish$(x)$, $\cos(x),\, \cos^2(x)$,\, $\tanh(x),\, \arctan(x)$. And we find an interesting and novel result that selecting the activation function according to the forms of solution and equation can usually achieve the better effects. On the other hand,
for inverse problems using data-driven models, the PINNs can be considered in the data-driven parameter discovery. For example, the parameter discovery of the potentials were found in the NLSE via the PINNs~\cite{Zhou-pla21}. However,
to the best of our knowledge, the inverse problems for $\PT$-symmetric potentials discovery rather than just the potential parameters were not fully discussed before. Therefore, the second novelty of our study is that, based on the PINNs deep learning framework, we present the modified PINNs (mPINNs) method  to identify the complex potential function, $V(\bx)+iW(\bx)$ included in the SNLSE (\ref{nls}) rather than just the potential parameters. In this way, we can determine the properties of the potential just by the solution. Especially, for the stationary solution of Eq.~(\ref{nls}) in the form of $\psi(\bx, z) = \phi(\bx)e^{i\mu z},\, \mu\in\mathbb{R}$, we can identify the complex potential of the SNLSE only by the PINNs employed in the corresponding stationary equation. Then two types of NN structures are compared under different parameter conditions.

The rest of this paper is arranged as follows. In Sec.~\ref{sec2}, we firstly introduce the PINNs deep learning framework for the 1D and 2D SNLSEs with $\PT$-symmetric potentials. And then the PINNs deep learning scheme is used to investigate the data-driven solitons and the general non-stationary solutions of the 1D and 2D SNLSEs (\ref{nls}) with two types of $\PT$ potentials ($\PT$  Scarf-II and periodic potentials). In Sec.~\ref{sec4}, some factors affecting the neural network performance are discussed in details including activation functions, structures of the networks, and the sizes of the training data. Moreover, several special activation functions are given to achieve the better effect. In Sec.~\ref{sec3}, we firstly propose the mPINNs method based on the PINNs deep learning framework to identify the potential function of the SNLSE rather than just the potential parameters in 1D and 2D cases. Then, for the stationary solution of Eq.~(\ref{nls}) in the form of $\psi(\bx, z) = \phi(\bx)e^{i\mu z}$, the PINNs can be used to identify the potential of the SNLSE. And the two types of network structures are compared under different parameter conditions. And the inverse problems about 1D and 2D $\PT$-symmetric potentials depending on propagation distance $z$ are also investigated using mPINNs method. Finally, some conclusions and discussions are presented in Sec.~\ref{sec5}.

\section{Data-driven solitons of the SNLSE with $\PT$ potentials}\label{sec2}

 In this section, we will firstly give the deep learning PINN framework for forward problems of the SNLSE (\ref{nls}) in Sec. 2.1. And then we use the scheme to successfully learn the solitons of the 1D and 2D SNLSEs with two fundamental $\PT$-symmetric potentials, namely, non-periodic Scarf-II potentials in Sec. 2.2 and periodic potentials in Sec. 2.3.

\subsection{The PINNs deep learning framework for forward problems of SNLSE (\ref{nls})}

 Firstly, we introduce the PINNs deep learning framework \cite{pinn} for the data-driven solutions of SNLSE (\ref{nls}). The main idea of the PINNs is to train a deep neural network constrained physical laws to fit the solutions of SNLSE (\ref{nls}). For the SNLSE (\ref{nls}) with the initial-boundary value conditions (notice that we will use $\psi(\bx,z=0)$ as the initial condition in the following)
\begin{equation}\label{ib}
 \left\{\begin{array}{l}
 \d i\psi_z+\nabla_{\bx}^2\psi+[V(\bx)+iW(\bx)]\psi+\frac{g|\psi|^2\psi}{1+S|\psi|^2}=0,\quad
(\bx, z)\in\Omega\times (0, Z),\v\\ \psi(\bx,0)=\phi_0(\bx),\quad\bx\in\Omega, \v\\
     \psi(\bx,z)\big|_{\bx\in \partial\Omega}=\phi_b(z),\quad z\in [0,Z],
 \end{array}\right.
\end{equation}
we rewrite the complex wave-function as $\psi(\bx,z)=p(\bx,z)+iq(\bx,z)$ with the real-valued function $p(\bx,z)$ and $q(\bx,z)$ being its real and imaginary parts, respectively. We now use a complex-valued deep neural network to approximate $\psi(x,z)$, and then,  based on the SNLE (\ref{nls}), the complex-valued PINNs $\mathcal{F}(\bx, z)$ is given by
\bee\label{F}
     \displaystyle\mathcal{F}(\bx, z):=-\mathcal{F}_p(\bx, z)+i\mathcal{F}_q(\bx, z)= i\hat{\psi}_z+\nabla_{\bx}^2\hat{\psi}+[V(\bx)+iW(\bx)]\hat{\psi}+\frac{g|\hat{\psi}|^2\hat{\psi}}{1+S|\hat{\psi}|^2}
\ene
with $-\mathcal{F}_p(\bx, z)$ and $\mathcal{F}_q(\bx, z)$ being its real and imaginary parts, respectively, defined as
   \begin{equation}\label{F1}
 \begin{array}{l}
     \displaystyle\mathcal{F}_p(\bx, z):= \hat{q}_z-\nabla^2_{\bx}\hat{p}-V(\bx)\hat{p}+W(\bx)\hat{q}-\frac{g(\hat{p}^2+\hat{q}^2)}{1+S(\hat{p}^2+\hat{q}^2)}\hat{p},\v\\
     \displaystyle\mathcal{F}_q(\bx, z):= \hat{p}_z+\nabla^2_{\bx}\hat{q}+V(\bx)\hat{q}+W(\bx)\hat{p}+\frac{g(\hat{p}^2+\hat{q}^2)}{1+S(\hat{p}^2+\hat{q}^2)}\hat{q}.
 \end{array}
\end{equation}
Furthermore, a Python library for PINNs, DeepXDE, was designed to serve a research tool for solving problems in computational science and engineering \cite{deepxde}.

\begin{figure*}[!t]
    \centering
  {\scalebox{0.8}[0.8]{\includegraphics{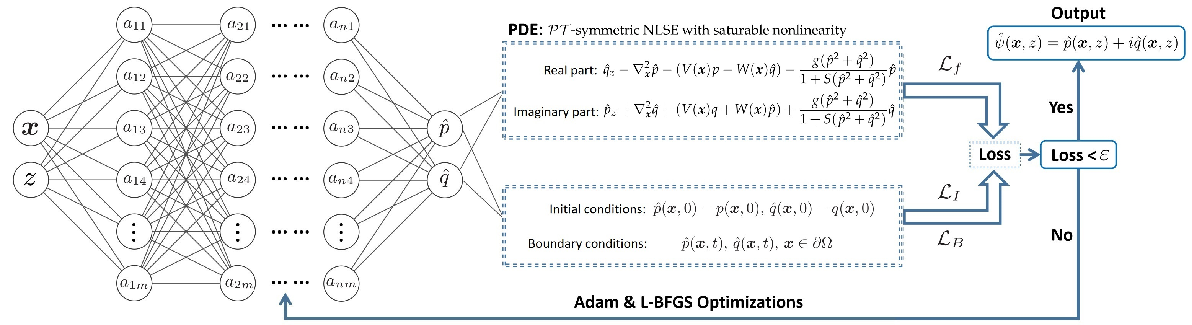}}}\hspace{-0.35in}
\vspace{0.2in}
\caption{\small The PINNs scheme solving the $\PT$-symmetric SNLSE with initial-boundary conditions (\ref{ib}).}
  \label{net1}
\end{figure*}


Therefore,  we can construct a fully-connected neural network NN$(\bx, z; W, B)$ with $n$ hidden layers and $m$ neurons in each layer to learn the hidden solution $ \hat\psi(x,z)=\hat p(x,z)+i \hat q(x,z)$ (see Fig.~\ref{net1}), where the parameters $W = \{w_j\}_{1}^{n+1}$ and $B = \{b_j\}_1^{n+1}$ being the weight matrices and bias vectors, respectively. Then  the vector data of the hidden layers and output layer can be generated by
\begin{equation}\label{sigma}
\begin{array}{l}
 A_j=\sigma(w_j\cdot A_{j-1}+b_j),\quad j=1,2,...,n,\quad
  A_{n+1}=w_{n+1}\cdot A_{n}+b_{n+1},
  \end{array}
\end{equation}
where $\sigma(\cdot)$ denotes some nonlinear activation function, $w_j$ is a dim$(A_j)\times $dim$(A_{j-1})$ matrix, $A_0$, $A_{n+1}$, $b_{n+1}\in \mathbb{R}^{2}$ and $A_j=(a_{j1},...,a_{jm})^T$, $b_j=(b_{j1},...,b_{jm})^T\in \mathbb{R}^m$.

To train the deep neural network to fit the solution of Eq.~(\ref{ib})  well, the total mean squared error (MSE) is used to define the loss function of the the neural network containing three parts
\begin{equation}\label{loss}
\begin{array}{rl}
  \mathcal{T\!L}=& \mathcal{L}_f+\mathcal{L}_I+\mathcal{L}_B, \v\\
  =& \d\frac{1}{N_f}\sum_{\ell=1}^{N_f}\left(|\mathcal{F}_p(\bx_f^\ell,z_f^\ell)|^2
     +|\mathcal{F}_q(\bx_f^\ell,z_f^\ell)|^2\right)+\frac{1}{N_I}\sum_{\ell=1}^{N_I}\left(|\hat{p}(\bx_I^\ell,0)-p_0^\ell|^2
     +|\hat{q}(\bx_I^\ell,0)-q_0^\ell|^2\right)\v\\
     &\d +\frac{1}{N_B}\sum_{\ell=1}^{N_B}\left(|\hat{p}(\bx_B^\ell,z_B^\ell)-p_B^\ell|^2
     +|\hat{q}(\bx_B^\ell,z_B^\ell)-q_B^\ell|^2\right),
 \end{array}
\end{equation}
where $\{\bx_f^\ell,z_f^\ell\}_\ell^{N_f}$ are connected with the randomly chosen sample points in $\Omega\times[0,Z]$ for the PINNs $\mathcal{F}(\bx, t)=-\mathcal{F}_p(\bx, t)+i\mathcal{F}_q(\bx, t)$,
$\{\bx_I^\ell,p_0^\ell,q_0^\ell\}_\ell^{N_I}$ represent the initial data with $\phi_0(\bx_I^\ell)=p_0^\ell+iq_0^\ell$, and $\{\bx_B^\ell,z_B^\ell,p_B^\ell,q_B^\ell\}_\ell^{N_B}$ are linked with the randomly selected boundary training data in domain $\partial\Omega\times[0,Z]$ with $\psi_b(\bx_B^\ell,z_B^\ell)=p_B^\ell+iq_B^\ell$. In addition to $\mathcal{L}_I$ and $\mathcal{L}_B$ making the solution satisfy the initial-boundary value conditions, the biggest advantage of PINNs is to introduce $\mathcal{L}_f$ as the loss to make the learning solution almost obey Eq.~(\ref{ib})  except that it satisfies the initial-boundary conditions. With the aid of some optimization approaches (e.g., Adam \& L-BFGS) \cite{adam,bfgs}, we minimize the whole MSE $\mathcal{T\!L}$ to make the approximated solution $\hat{\psi}=\hat{p}+i\hat{q}$ satisfy Eq.~(\ref{ib}) (see Fig.~\ref{net1} for the PINNs scheme in detail).

We here choose a hyperbolic tangent function $\tanh(\cdot)$ as the activation function (of course one can also choose other nonlinear functions as the activation functions, e.g., Sigmoid, ReLU, leaky ReLU, sinc, ELU, softmax, Swish functions or some other nonlinear functions, see Sec. 3 for the detailed comparisons), and use Glorot normal to initialize variate.
The main steps of the PINNs method solving the $\PT$-symmetric SNLSE (\ref{ib}) with initial-boundary value conditions are
presented in Table~\ref{table-pinns}.
\begin{table}[!t]
\centering
\caption{The PINNs method learning the $\PT$-symmetric SNLSE (\ref{ib}).}
\begin{tabularx}{\textwidth}{lX}
\hline\hline
Step &   Instruction \\
\hline
1 &  Establishing a fully-connected neural network NN$(\bx, z; W, B)$ with initialized parameters $W = \{w_j\}_{1}^{n+1}$ and $B = \{b_j\}_1^{n+1}$ being the weights and bias, respectively, and the PINNs $\mathcal{F}(\bx, z)$ is given by Eq.~(\ref{F}), and choosing the nonlinear activation function (see  Eq.~(\ref{sigma}));  \\[4ex]
2 & Generating three training data sets for the initial-boundary value conditions and considered physical model (\ref{F}), respectively, from the initial-boundary and region;  \\[1ex]
 3& Constructing a training loss function $\mathcal{T\!L}$ given by Eq.~(\ref{loss}) by summing the MSE of both the $\mathcal{F}(\bx, z)$ and initial-boundary value residuals; \\[1ex]
4 & Training the NN to optimize the parameters $\{W, B\}$ by minimizing the loss function $\mathcal{TL}$ in terms of
the Adam \& L-BFGS optimization algorithm.  \\[1ex]
\hline\hline
\end{tabularx}
\label{table-pinns}
\end{table}


\subsection{Data-driven solitons of the SNLSE with $\PT$ Scarf-II potential}

In this subsection, we will use the schemes to successfully learn the solitons of the 1D and 2D SNLSEs with
$\PT$-symmetric non-periodic Scarf-II potentials.

\subsubsection{1D SNLSE with $\PT$ Scarf-II potential}

Firstly, we consider the 1D $\PT$-symmetric Scarf-II potential~\cite{Ah01}
\bee \label{scarf}
 V(x)=V_0\,{\rm sech}^2x, \quad
 W(x)=W_0\,{\rm sech} x\tanh x,
\ene
with the real-valued parameters $V_0$ and $W_0$ being the amplitudes or strengths of external potential (real part) and gain-and-loss distribution (imaginary part) of the $\PT$ Scarf-II potential, respectively. The stationary solution to Eq.~(\ref{nls}) is sought in the form
$\psi(x,z)=\phi(x)e^{i\mu z}$, where $\mu$ represents the real-valued propagation constant, and  $\phi(x)\in \mathbb{C}[x]$ with $\lim_{|x|\rightarrow \infty}\phi(x)=0$ satisfies the nonlinear stationary equation
\begin{equation}\label{nls-s}
   \mu\phi=\phi_{xx}+[V(x)+iW(x)]\phi+\frac{g|\phi|^2\phi}{1+S|\phi|^2}.
\end{equation}
Since the self-focusing ($g=1$) and defocusing ($g=-1$) cases are similar, we shall consider the self-focusing case  $g=1$ hereafter, and always take $S=1$.

 To generate the training data, we firstly utilize Newton conjugate-gradient method \cite{NCG} to obtain a numerical soliton $\phi(x)$  of Eq.~(\ref{nls-s}) with the zero-boundary condition and $\mu=1,\, V_0=1,\, W_0=0.5$, and further use it as the initial condition to generate a .mat data-set about $\psi(x, z)$ in the domain $\Omega \times [0, Z]$ with the 256 Fourier modes in the $x$ direction and propagation-step $\Delta z = 0.01$ by the Fourier spectral method in Matlab \cite{spectral}. As a result, we generate the initial data and the `exact' data in the domain for the study of deep PINNs.

\begin{figure*}[!t]
    \centering
\vspace{-0.15in}
  {\scalebox{0.45}[0.45]{\includegraphics{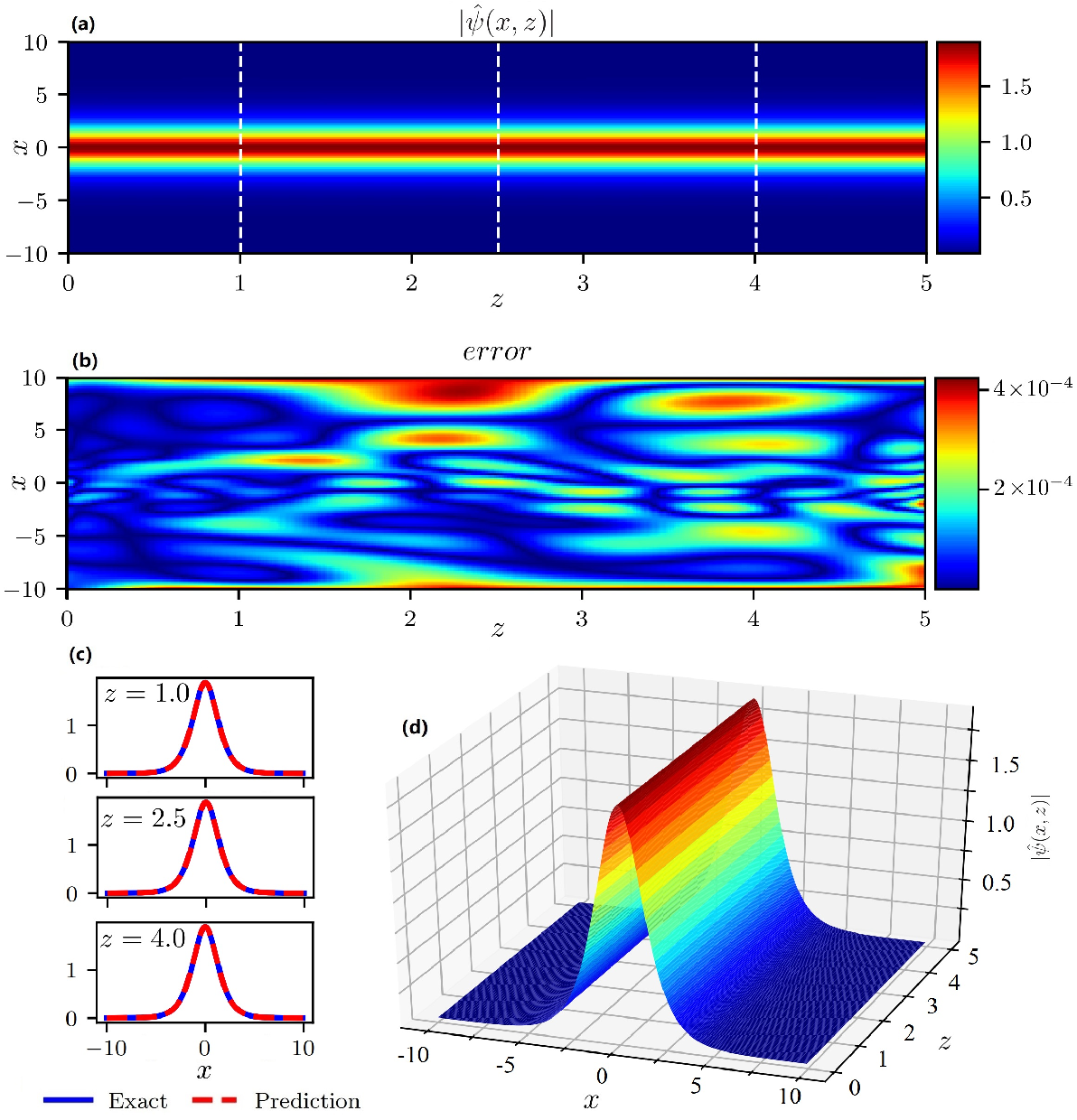}}}\hspace{-0.35in}
\vspace{0.1in}
\caption{\small Data-driven solitons of 1D SNLSE with $\PT$ Scarf-II potential (\ref{scarf}): (a) 2D profile of the learning soliton solution; (b) The module of absolute error between the exact and learning solutions, $error=|\hat{\psi}-\psi|$; (c) The soliton solutions at different propagation distance $z = 1.0,\, 2.5$, and $4.0$; (d) The 3D profiles of the learning soliton solution.}
  \label{fig1}
\end{figure*}

Here we take the potential parameters as $V_0=1$ and $W_0=0.5$, and consider $\Omega=[-10,10]$ and $Z=5$ in Eq.~(\ref{ib}). We choose a 4-layer deep neural network with 100 neurons per layer, and take the random sample points $N_f=2000$, $N_B=50$ and $N_I=100$, respectively. Then, by using 5000 steps Adam and 5000 steps L-BFGS optimizations, we obtain the learning soliton solution $\hat{\psi}(x,z)$, whose 2D and 3D profiles are shown in Figs.~\ref{fig1}(a, d). And the module of absolute error between of exact (obtained by the numerical method in Matlab) and learning solutions $error=|\hat{\psi}-\psi|$ is also displayed in Fig.~\ref{fig1}(b). Besides the comparisons of predicted solution (red dashed line) and exact solution (blue solid line) are presented at three different propagation distance $z = 1.0,\, 2.5$, and $4.0$, which show that the predicted solutions match well with the exact ones at different locations (see Fig.~\ref{fig1}(c)). The relative $\mathbb{L}^2$ norm errors of $\psi(x, z)$, $p(x, z)$ and $q(x, z)$, respectively, are $2.337\cdot 10^{-4}$, $3.491\cdot 10^{-4}$ and $3.209\cdot 10^{-4}$. Finally, we should mention that the learning times of Adam and L-BFGS optimizations are 185s and 222s, respectively, by using a Lenovo notebook with a 2.30GHz eight-cores i7 processor and a RTX3080 graphics processor.

\begin{figure*}[!h]
    \centering
\vspace{-0.15in}
  {\scalebox{0.5}[0.5]{\includegraphics{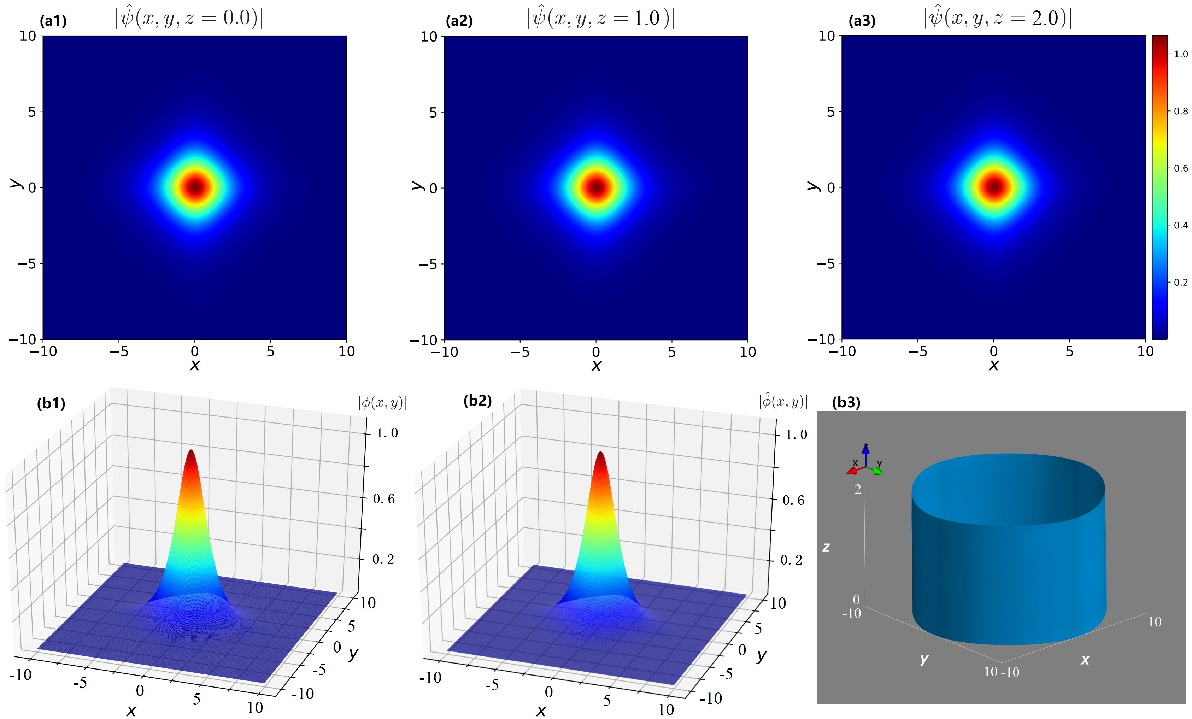}}}\hspace{-0.35in}
\vspace{0.15in}
\caption{\small Data-driven solitons of 2D SNLSE with $\PT$ Scarf-II potential (\ref{scarf2D}): (a1, a2, a3) The magnitude of the learning solutions at different propagation distances $z = 0,\, 1.0$, and $2.0$, respectively. (b1, b2) The initial state ($\phi(x,y)=\psi(x,y,z)=0)$) of the exact and predicted solitons. (b3) The contour lines of learning soliton intensity.}
  \label{fig2}
\end{figure*}

\subsubsection{2D SNLSE with $\PT $Scarf-II potential}

Similarly, we consider data-driven solitons of the 2D SNLSE in the $\PT$ Scarf-II potential
\begin{equation}\label{scarf2D}
 \begin{array}{l}
     V(x,y)=V_0({\rm sech}^2x+{\rm sech}^2y),\quad
     W(x,y)=W_0({\rm sech} x\tanh x+{\rm sech} y\tanh y),
 \end{array}
\end{equation}
where the real-valued parameters $V_0$ and $W_0$ control the amplitudes of the real and imaginary parts of the $\PT$ potential, respectively. Analogously, the stationary solution of the 2D NLSE with SN and  $\PT$ Scarf-II potential (\ref{scarf2D}) can be given by $\psi(x,y,z)=\phi(x,y)e^{i\mu z}$ with $\phi(x,y)\in \mathbb{C}[x,y]$ and $\lim_{|\bx|\rightarrow \infty}\phi(\bx)=0$ obeying
\begin{equation}\label{nls-s2}
   \mu\phi=\phi_{xx}+\phi_{yy}+[V(x,y)+iW(x,y)]\phi+\frac{g|\phi|^2\phi}{1+S|\phi|^2}.
\end{equation}
  And $\phi(x,y)$ of Eq.~(\ref{nls-s2}) can be obtained by Newton conjugate-gradient method with $\mu=1$. Meanwhile, we generate a .mat high-accuracy data-set about $\psi(x, y, z)$ in the domain $\Omega \times [0, Z]$ with the 512 Fourier modes in the $x$ direction and propagation-step $\Delta z = 0.005$, which is used as an error check.

\begin{figure*}[!t]
    \centering
\vspace{-0.15in}
  {\scalebox{0.45}[0.46]{\includegraphics{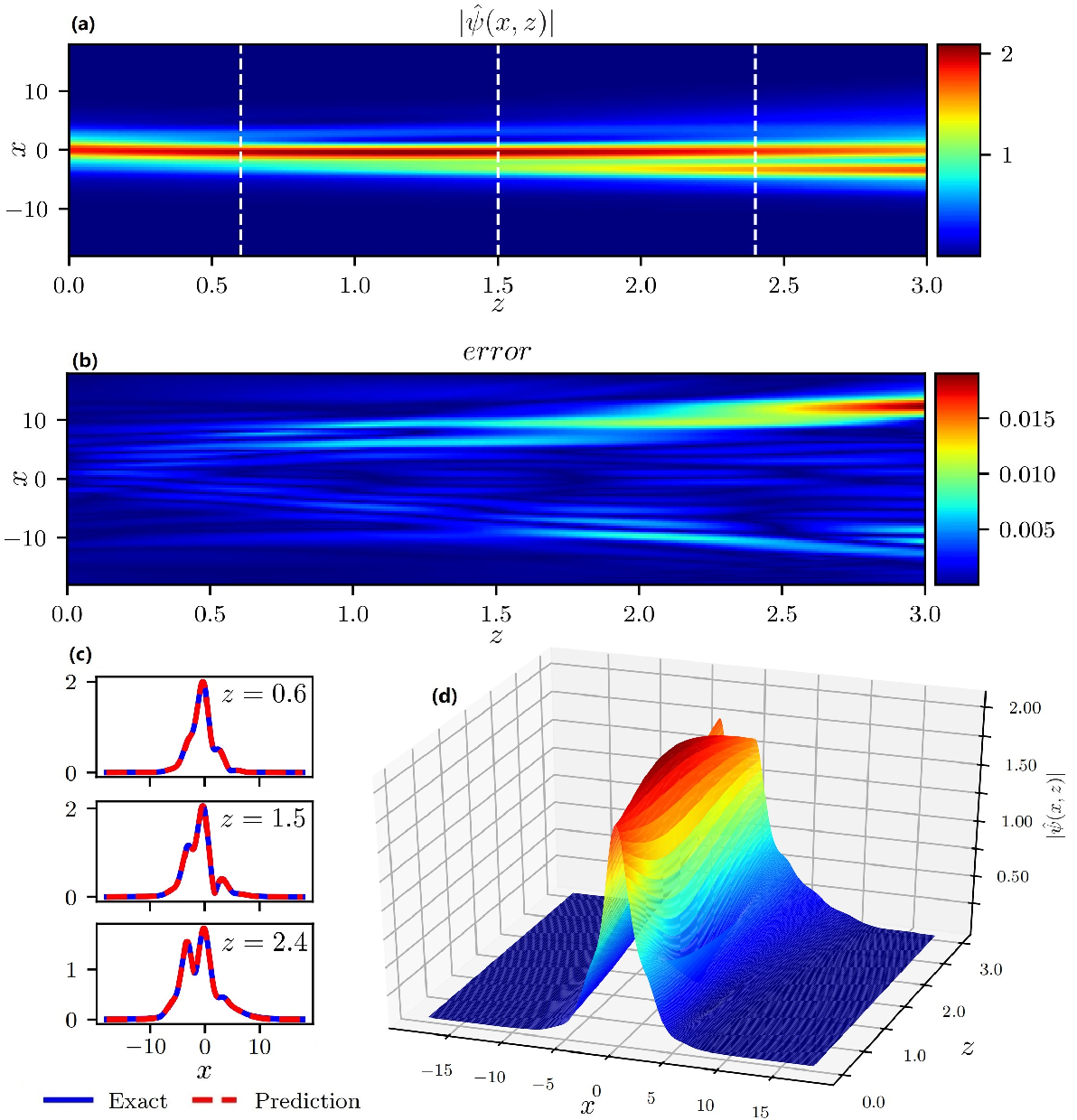}}}\hspace{-0.35in}
\vspace{0.1in}
\caption{\small Data-driven non-stationary solution of 1D SNLSE with $\PT$ periodic potential (\ref{period}): (a) 2D profile of the learning solution; (b) The module of absolute error between the exact and learning solutions $error=|\hat{\psi}-\psi|$; (c) The non-stationary solutions at different propagation distances $z = 0.6,\, 1.5$, and $2.4$; (d) 3D profile of the learning non-stationary solution.}
  \label{fig3}
\end{figure*}

Particularly, we choose $V_0=1$, $W_0=0.5$ and $(x,y)\in\Omega=[-10,10]\times[-10,10]$, $Z=2$, and consider a 4-hidden-layer deep neural network with 100 neurons per layer. And we randomly choose points in domain $\Omega\times[0,Z]$, where $N_f=20000$, $N_B=200$ and $N_I=2000$, respectively. Then, by using 15000 steps Adam and 49 steps L-BFGS optimizations, the predicted solution $\hat{\psi}(x,y,z)$ is obtained. Figs.~\ref{fig2}(a1, a2, a3) exhibit the magnitudes of the predicted solution at different propagation distances $z = 0,\, 1.0$, and $2.0$, respectively. And the initial states ($\phi(x,y)=\psi(x,y,z)=0)$) of the exact and predicted solitons are shown in Figs.~\ref{fig2}(b1, b2), respectively. Furthermore, nonlinear propagation simulation of the learning 2D soliton is displayed by contour lines of soliton intensity in Fig.~\ref{fig2}(b3), with the value of the contour lines taken as $\max\left(\|\hat{\psi}(x,y,0)\|_2\right)/2$ hereinafter, which reveals that the stationary soliton is stable in a short distance propagation. The relative $\mathbb{L}^2$ norm errors of $\psi(x, y, z)$, $p(x, y, z)$ and $q(x, y, z)$, respectively, are $6.131\cdot 10^{-3}$, $7.593\cdot 10^{-3}$ and $8.136\cdot 10^{-3}$. And the learning times of Adam and L-BFGS optimizations are 675s and 2s, respectively.

We should mention that the training stops in each step of the L-BFGS optimization when
\begin{equation}\label{stop}
  \frac{L_k-L_{k+1}}{\max\{|L_k|,|L_{k+1}|,1\}}\leq {\rm np.finfo(float).eps},
\end{equation}
where $L_k$ denotes loss in the n-th step L-BFGS optimization, and np.finfo(float).eps represent Machine Epsilon. Here we always set the default float type to `float64'. When the relative error between $L_k$ and $L_{k+1}$ is less than Machine Epsilon, the iteration stops. Therefore, in the previous training only 49 steps L-BFGS optimization are performed owing to the relative error have been less than Machine Epsilon.

\begin{figure*}[!t]
    \centering
\vspace{-0.15in}
  {\scalebox{0.56}[0.56]{\includegraphics{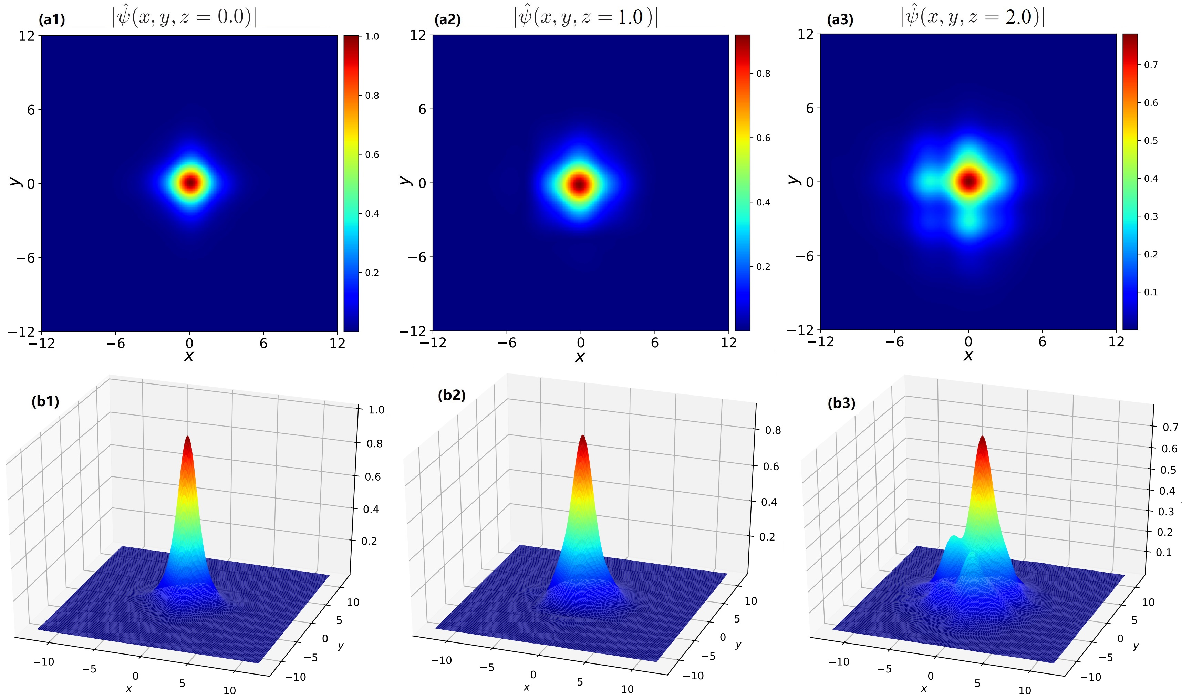}}}\hspace{-0.35in}
\vspace{0.1in}
\caption{\small Data-driven non-stationary solution of 2D SNLSE with $\PT$ periodic potential (\ref{period2D}): (a1, a2, a3) The magnitude of the learning solution at different propagation distance $z = 0,\, 1.0$, and $2.0$, respectively. (b1, b2, b3) The corresponding 3D profiles at different propagation distance $z = 0,\, 1.0$, and $2.0$, respectively.}
  \label{fig4}
\end{figure*}

\subsection{Data-driven solitons of the SNLSEs with $\PT$ periodic potentials}

In this subsection, we will consider the data-driven solitons of the 1D and 2D SNLSE with $\PT$ periodic potentials via the PINNs.

\subsubsection{1D SNLSE with $\PT$ periodic potential}

We investigate the solutions of 1D SNLSE with $\PT$ periodic potential (alias lattice potential)~\cite{pt1}
\bee \label{period}
 V(x)=\cos^2(x), \qquad
 W(x)=W_0\,\sin(2x),
\ene
where $W_0$ controls the gain-and-loss distribution of the optical potential. In general, the instability growth rate of the solution tends to increase with $W_0$.

For the 1D SNLSE (\ref{nls}) with the $\PT$ periodic potential \eqref{period}, we consider its non-stationary solutions. First, we use the numerical method to find the stationary soliton $\phi(x)$ of the corresponding stationary equation (\ref{nls-s}) with the $\PT$ periodic potential \eqref{period}. Then we use the function $\tilde{\phi}(x)=\phi(x)e^{-i0.5x}$ (which dose not satisfy the stationary equation (\ref{nls-s})) as the initial condition to generate a .mat data-set about $\psi(x,z)$ of the 1D SNLSE (\ref{nls}) with the $\PT$ periodic potential \eqref{period} in the domain $\Omega\times [0,Z]$ with the 512 Fourier modes in the $x$ direction and propagation-step $\Delta z = 0.005$ by the Fourier spectral method. As a result, we generate the initial data and the `exact' data in the domain for the study of deep PINNs.

Here we take $W_0=0.5$,\, $\Omega=[-18,18]$ and $Z=3$, and choose a 4-hidden-layer deep neural network with 100 neurons per layer and random sample points $N_f=4000$, $N_B=50$ and $N_I=100$, respectively. Then after the 5000 steps Adam and 10000 steps L-BFGS optimizations, we can obtain the learning non-stationary solution, whose two-dimensional and three-dimensional profiles are exhibited in Figs.~\ref{fig3}(a, d). And the module of absolute error between the exact and learning solutions $error=|\hat{\psi}-\psi|$ is also calculated (see Fig.~\ref{fig3}(b)). Furthermore, the comparison of predicted solution (red dashed line) and exact solution (blue solid line) is displayed at different propagation distance $z = 0.6,\, 1.5$, and $2.4$ which shows that the solution is unstable in propagation (see Fig.~\ref{fig3}(c)). The relative $\mathbb{L}^2$ norm errors of $\psi(x, z)$, $p(x, z)$ and $q(x, z)$, respectively, are $5.400\cdot 10^{-3}$, $7.719\cdot 10^{-3}$ and $6.739\cdot 10^{-3}$. Finally, we should mention that the learning times of Adam and L-BFGS optimizations are 270s and 617s, respectively.

\subsubsection{2D SNLSE with $\PT$ periodic potential}

Similarly, we here discuss the non-stationary solutions in 2D $\PT$ periodic potential~\cite{pt1}
\bee \label{period2D}
 \begin{array}{l}
     V(x,y)=\cos^2(x)+\cos^2(y),\quad
     W(x,y)=W_0[\sin(2x)+\sin(2y)],
 \end{array}
\ene
where $W_0$ controls the amplitude of the gain-and-loss distribution.

 Here we take the function $\phi(x,y)={\rm sech}(x){\rm sech}(y)$ as the initial condition, which does not satisfy the stationary equation (\ref{nls-s2}), to generate a .mat data-set about $\psi(x,y,z)$ of the 2D SNLSE (\ref{nls}) with the $\PT$ periodic potential \eqref{period2D} at $W_0=0.5$ in the domain $\Omega\times [0,Z]$ with the 512 Fourier modes in the $x$ and $y$ direction and propagation-step $\Delta z = 0.005$ by the Fourier spectral method. Especially, we choose $W_0=0.5$ and $\Omega=[-12,12]\times[-12,12]$, $Z=2$.  And we take a 4-hidden-layer deep neural network with 32 neurons per layer, random sample points $N_f=30000$, $N_B=100$ and $N_I=1000$, respectively. Then, by using 15000 steps Adam and 25000 steps L-BFGS optimizations, we can obtain the predicted solution $\hat{\psi}(x,y,z)$. Figs.~\ref{fig4}(a1, a2, a3) exhibit the magnitudes of the predicted solution at different propagation distances $z = 0,\, 1.0$, and $2.0$, respectively. And the corresponding 3D profiles are shown in Figs.~\ref{fig4}(b1, b2, b3), respectively. It is obvious that the amplitude and shape of the solution change in propagation which reveals that it is non-stationary. The relative $\mathbb{L}^2$ norm errors of $\psi(x, y, z)$, $p(x, y, z)$ and $q(x, y, z)$, respectively, are $1.737\cdot 10^{-2}$, $2.973\cdot 10^{-2}$ and $3.087\cdot 10^{-2}$. And the learning times of Adam and L-BFGS optimizations are 1259s and 2135s, respectively.

\begin{figure*}[!t]
    \centering
  {\scalebox{0.85}[0.85]{\includegraphics{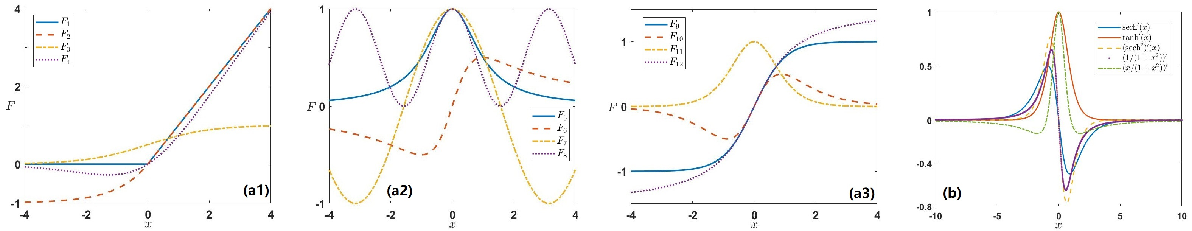}}}
\vspace{0.15in}
\caption{\small (a1, a2, a3) Profiles of 12 nonlinear activation functions: $F_1={\rm ReLU}(x)$, $F_2={\rm ELU}(x)$, $F_3={\rm Sigmoid}(x)$, $F_4={\rm Swish}(x)$, $F_5=1/(1+x^2)$,\, $F_6=x/(1+x^2)$, $F_7=\cos(x)$,\, $F_8=\cos^2(x)$,\, $F_9=\tanh(x)$,\,
$F_{10}={\rm sech}(x)\tanh(x),$\, $F_{11}={\rm sech}^2(x),\, F_{12}=\arctan(x)$.  (b) Profiles of the derivatives of activation functions $\mathrm{sech}(x)$, $F_5$, $F_6$, $F_9$ and $F_{11}$.} 
  \label{act}
\end{figure*}
\begin{table}[!h]
\begin{center}\small
\caption{\small $\mathbb{L}^2$-norm errors of $p(\bx,z)$, $q(\bx,z)$ and $\psi(\bx,z)$ of the SNLSE with 1D/2D $\PT$ Scarf-II or periodic potential for 12 different activation functions, where $\mathcal{Z} = w_j\cdot A_{j-1} + b_j$ with $w_j$ and $b_j$ being weights and bias, respectively.}\label{table3_1}
\vspace{0.05in}
\begin{tabular}{ cccccccc }
\hline
\hline\\[-2ex]
 &  &  \multicolumn{6}{c}{Type of nonlinear activation functions}  \\[0.2ex]
\cline{3-8}\\[-2ex]
$\PT$ potentials& errors & ReLU$(\mathcal{Z})$            & ELU$(\mathcal{Z})$           & Sigmoid$(\mathcal{Z})$     & Swish$(\mathcal{Z})$       & 1/$(1+\mathcal{Z}^2)$     &  $\mathcal{Z}/(1+\mathcal{Z}^2)$ \\[0.5ex]
\hline

\rule{0pt}{13pt}                &$p$         & $8.919 \cdot 10^{-1}$    & $7.395 \cdot 10^{-1}$  & $2.418 \cdot 10^{-2}$& $3.284 \cdot 10^{-4}$& $7.563 \cdot 10^{-4}$&  $3.286 \cdot 10^{-4}$\\
\rule{0pt}{13pt}  1D Scarf-II   &$q$         & $9.592 \cdot 10^{-1}$    & $8.827 \cdot 10^{-1}$  & $2.454 \cdot 10^{-2}$& $4.282 \cdot 10^{-4}$& $7.041 \cdot 10^{-4}$&  $3.849 \cdot 10^{-4}$\\
\rule{0pt}{13pt}               &$\psi$       & $7.217 \cdot 10^{-1}$    & $6.256 \cdot 10^{-1}$  & $1.648 \cdot 10^{-2}$& $3.088 \cdot 10^{-4}$& $5.046 \cdot 10^{-4}$&  $2.302 \cdot 10^{-4}$\\[0.5ex]
\hline

\rule{0pt}{13pt}                &$p$         & $5.981 \cdot 10^{-1}$    & $2.006 \cdot 10^{-1}$  & $1.288 \cdot 10^{-1}$& $7.254 \cdot 10^{-3}$& $7.415 \cdot 10^{-3}$&  $2.083 \cdot 10^{-2}$\\
\rule{0pt}{13pt}  2D Scarf-II   &$q$         & $6.569 \cdot 10^{-1}$    & $2.328 \cdot 10^{-1}$  & $1.427 \cdot 10^{-1}$& $6.007 \cdot 10^{-3}$& $6.564 \cdot 10^{-3}$&  $1.981 \cdot 10^{-2}$\\
\rule{0pt}{13pt}               &$\psi$       & $5.914 \cdot 10^{-1}$    & $1.977 \cdot 10^{-1}$  & $1.114 \cdot 10^{-1}$& $4.923 \cdot 10^{-3}$& $4.891 \cdot 10^{-3}$&  $1.503 \cdot 10^{-2}$\\[0.5ex]
\hline

\rule{0pt}{13pt}                &$p$         & $7.532 \cdot 10^{-1}$    & $4.569 \cdot 10^{-1}$  & $5.410 \cdot 10^{-2}$& $8.460 \cdot 10^{-3}$& $6.956 \cdot 10^{-3}$&  $6.964 \cdot 10^{-3}$\\
\rule{0pt}{13pt}   1D Periodic  &$q$         & $7.271 \cdot 10^{-1}$    & $4.605 \cdot 10^{-1}$  & $6.502 \cdot 10^{-2}$& $6.507 \cdot 10^{-3}$& $5.430 \cdot 10^{-3}$&  $5.059 \cdot 10^{-3}$\\
\rule{0pt}{13pt}               &$\psi$       & $6.498 \cdot 10^{-1}$    & $3.790 \cdot 10^{-1}$  & $4.919 \cdot 10^{-2}$& $5.937 \cdot 10^{-3}$& $5.109 \cdot 10^{-3}$&  $5.091 \cdot 10^{-3}$\\[0.5ex]
\hline

\rule{0pt}{13pt}                &$p$         & $8.972 \cdot 10^{-1}$    & $4.030 \cdot 10^{-1}$  & $7.752 \cdot 10^{-2}$& $3.986 \cdot 10^{-2}$& $3.894 \cdot 10^{-2}$&  $5.592 \cdot 10^{-2}$\\
\rule{0pt}{13pt}   2D Periodic  &$q$         & $8.584 \cdot 10^{-1}$    & $4.725 \cdot 10^{-1}$  & $7.792 \cdot 10^{-2}$& $4.269 \cdot 10^{-2}$& $4.059 \cdot 10^{-2}$&  $5.775 \cdot 10^{-2}$\\
\rule{0pt}{13pt}               &$\psi$       & $8.250 \cdot 10^{-1}$    & $4.108 \cdot 10^{-1}$  & $6.001 \cdot 10^{-3}$& $3.134 \cdot 10^{-2}$& $3.225 \cdot 10^{-2}$&  $4.394 \cdot 10^{-2}$\\[0.5ex]
\hline
\hline\\[-2ex]
 &  &  \multicolumn{6}{c}{Type of nonlinear activation functions}  \\[0.2ex]
\cline{3-8}\\[-2ex]
$\PT$ potentials& errors & cos$(\mathcal{Z})$             & $\cos^2(\mathcal{Z})$        & tanh$(\mathcal{Z})$        & sech$(\mathcal{Z})$tanh$(\mathcal{Z})$  & ${\rm sech}^2(\mathcal{Z})$ & arctan$(\mathcal{Z})$  \\[0.5ex]
\hline

\rule{0pt}{13pt}                &$p$         & $6.300 \cdot 10^{-4}$    & $7.432 \cdot 10^{-4}$  & $4.896 \cdot 10^{-4}$& $3.007 \cdot 10^{-4}$& $3.805 \cdot 10^{-4}$&  $6.965 \cdot 10^{-4}$\\
\rule{0pt}{13pt}  1D Scarf-II   &$q$         & $5.992 \cdot 10^{-4}$    & $6.843 \cdot 10^{-4}$  & $4.530 \cdot 10^{-4}$& $2.648 \cdot 10^{-4}$& $3.405 \cdot 10^{-4}$&  $7.346 \cdot 10^{-4}$\\
\rule{0pt}{13pt}               &$\psi$       & $4.364 \cdot 10^{-4}$    & $4.229 \cdot 10^{-4}$  & $2.946 \cdot 10^{-4}$& $2.244 \cdot 10^{-4}$& $2.274 \cdot 10^{-4}$&  $5.134 \cdot 10^{-4}$\\[0.5ex]
\hline

\rule{0pt}{13pt}                &$p$         & $2.045 \cdot 10^{-2}$    & $1.566 \cdot 10^{-2}$  & $8.135 \cdot 10^{-3}$& $3.877 \cdot 10^{-3}$& $2.539 \cdot 10^{-3}$&  $2.027 \cdot 10^{-2}$\\
\rule{0pt}{13pt}  2D Scarf-II   &$q$         & $1.841 \cdot 10^{-2}$    & $1.635 \cdot 10^{-2}$  & $7.987 \cdot 10^{-3}$& $3.632 \cdot 10^{-3}$& $2.404 \cdot 10^{-3}$&  $2.147 \cdot 10^{-2}$\\
\rule{0pt}{13pt}               &$\psi$       & $1.566 \cdot 10^{-2}$    & $1.252 \cdot 10^{-2}$  & $6.122 \cdot 10^{-3}$& $2.826 \cdot 10^{-3}$& $1.918 \cdot 10^{-3}$&  $1.601 \cdot 10^{-2}$\\[0.5ex]
\hline

\rule{0pt}{13pt}                &$p$         & $1.932 \cdot 10^{-3}$    & $1.671 \cdot 10^{-3}$  & $8.204 \cdot 10^{-3}$& $5.578 \cdot 10^{-3}$& $6.977 \cdot 10^{-3}$&  $8.668 \cdot 10^{-3}$\\
\rule{0pt}{13pt}   1D Periodic  &$q$         & $1.800 \cdot 10^{-3}$    & $1.519 \cdot 10^{-3}$  & $6.616 \cdot 10^{-3}$& $4.938 \cdot 10^{-3}$& $5.760 \cdot 10^{-3}$&  $7.174 \cdot 10^{-3}$\\
\rule{0pt}{13pt}               &$\psi$       & $1.264 \cdot 10^{-3}$    & $1.006 \cdot 10^{-3}$  & $5.677 \cdot 10^{-3}$& $3.858 \cdot 10^{-3}$& $4.751 \cdot 10^{-3}$&  $5.782 \cdot 10^{-3}$\\[0.5ex]
\hline

\rule{0pt}{13pt}                &$p$         & $3.733 \cdot 10^{-2}$    & $1.612 \cdot 10^{-2}$  & $6.140 \cdot 10^{-2}$& $4.821 \cdot 10^{-2}$& $4.581 \cdot 10^{-2}$&  $5.906 \cdot 10^{-2}$\\
\rule{0pt}{13pt}   2D Periodic  &$q$         & $3.712 \cdot 10^{-2}$    & $1.114 \cdot 10^{-2}$  & $6.343 \cdot 10^{-2}$& $4.935 \cdot 10^{-2}$& $4.764 \cdot 10^{-2}$&  $6.295\cdot 10^{-2}$\\
\rule{0pt}{13pt}               &$\psi$       & $2.908 \cdot 10^{-2}$    & $9.984 \cdot 10^{-3}$  & $4.768 \cdot 10^{-2}$& $3.663 \cdot 10^{-2}$& $3.615 \cdot 10^{-2}$&  $4.623 \cdot 10^{-2}$\\[0.5ex]
\hline
\hline
\end{tabular}
\end{center}
\end{table}

\begin{figure}[!t]
    \centering
\vspace{-0.15in}
  {\scalebox{0.75}[0.7]{\includegraphics{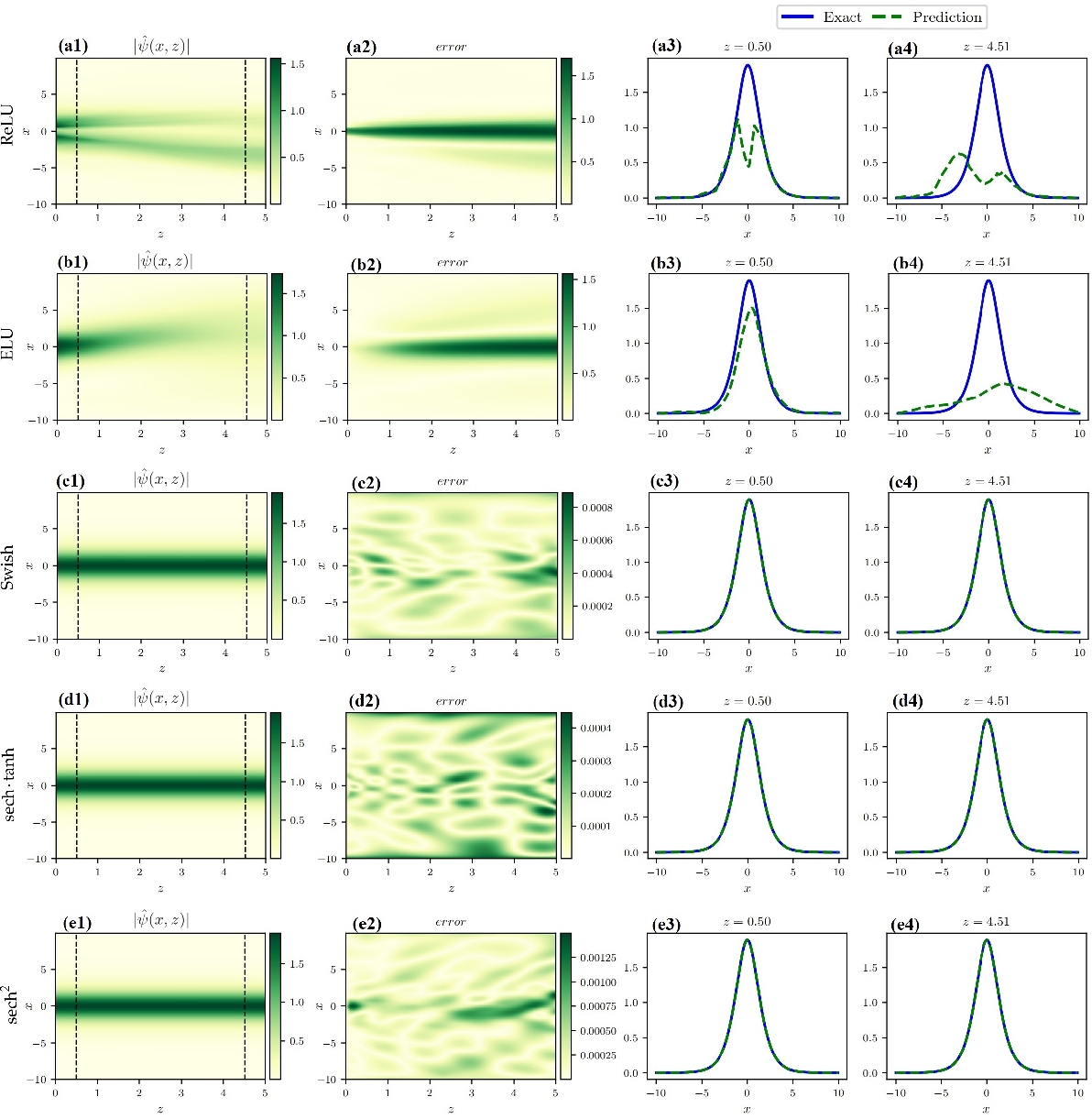}}}\hspace{-0.35in}
\vspace{0.1in}
\caption{\small The results for 1D Scarf-II potential with different activation functions. Rows one to five, respectively, represent the different activation functions: ReLU, ELU, Swish, sech$\cdot$tanh, and sech$^2$. Columns one to four, respectively, represent the predicted magnitudes of solutions, error values in magnitudes of solutions, and solitons at different propagation distances $z = 0.5$ and $4.51$.}
  \label{scarf1D_act}
\end{figure}

\section{Some main factors affecting the deep neural network performance}\label{sec4}

In this section, we study some main factors affecting the neural network performance, including activation functions, structures of the networks, and the sizes of the training data. Moreover, several special and new activation functions are shown to achieve the better effect.

\subsection{Impacts of nonlinear activation functions}

Firstly, we discuss the impacts of different nonlinear activation functions on the neural network performance. To consider the 1D and 2D $\PT$-symmetric Scarf-II and periodic potentials, we choose 12 special classes of activation functions (see Fig.~\ref{act})
\begin{equation}\label{a1}
  F_1={\rm ReLU}(x)=\max(0,x),\quad
  F_2= {\rm ELU}(x)=\left\{\begin{array}{ll}
 x, & x>0,\v\\
e^x-1, & x\leq0,
 \end{array}\right.
\end{equation}
\begin{equation}\label{a2}
  F_3={\rm Sigmoid}(x)=\frac{1}{1+e^{-x}}, \quad F_4={\rm Swish}(x)=\frac{x}{1+e^{-x}},
\end{equation}
\begin{equation}\label{a3}
  F_5=\frac{1}{1+x^2},\quad F_6=\frac{x}{1+x^2}, \quad F_7=\cos(x),\quad F_8=\cos^2(x),
\end{equation}
\begin{equation}\label{a4}
  F_9=\tanh(x),\quad F_{10}={\rm sech}(x)\tanh(x),\quad F_{11}={\rm sech}^2(x),\quad  F_{12}=\arctan(x).
\end{equation}
to find the data-driven solutions of the SNLSE with $\PT$ Scarf-II or periodic potential in 1D and 2D geometries. And it is concluded that selecting the activation functions according to the structures of solutions of equations usually can achieve the better effect.

Furthermore, we analyze the variance of the gradient of the activation value from the perspective of forward and backward propagations. For example, for weight $W=\{w_i\}_1^{n+1}$, its update iteration satisfies the following formula
\begin{equation}
  W_{k+1}=W_k-\eta\frac{\partial\mathcal{T\!L}}{\partial W},
\end{equation}
$\eta$ is learning rate and $\mathcal{T\!L}$ is loss function defined in Eq.~(\ref{loss}). By the chain rule, it can be obtained from Eq.~(\ref{sigma})
\begin{equation}
  \frac{\partial\mathcal{T\!L}}{\partial w_i}=\frac{\partial\mathcal{T\!L}}{\partial A_{n+1}}\frac{\partial A_{n+1}}{\partial A_{n}}\cdots \frac{\partial A_{i}}{\partial w_i},
\end{equation}
\begin{equation}
  \frac{\partial A_{n+1}}{\partial A_{n}}=w_{n+1},\quad \frac{\partial A_{i}}{\partial w_i}=\sigma'(w_i\cdot A_{i-1}+b_i)A_{i-1},
\end{equation}
\begin{equation}
  \frac{\partial A_{k}}{\partial A_{k-1}}=\sigma'(w_k\cdot A_{k-1}+b_k)w_k, \, k=i-1,\ldots n.
\end{equation}
Therefore, $\frac{\partial\mathcal{T\!L}}{\partial w_i}$ involves the powers of the derivative of the activation function and weight.
Meanwhile, the derivatives of these activation functions are exhibited in Fig.~\ref{act}(b), whose values are between $0.5$ and $1$. Considering the weight initialization method (Glorot normal), weight  and bias follow a normal distribution $N(0,\sigma^2)$, where \begin{equation}
 \sigma^2=\frac{2}{w_{\mathrm{in}}+w_{\mathrm{out}}}
\end{equation}
with $w_{\mathrm{in}}$ and $w_{\mathrm{out}}$ being the numbers of input and output units in the tensor, respectively. Besides, the used network layers are small. Therefore, the powers of the derivative of the activation function and weight will not go to zero or infinity. In summary, there will be no gradient disappearance and gradient explosion.

\begin{figure*}[!t]
    \centering
\vspace{-0.15in}
  {\scalebox{0.75}[0.75]{\includegraphics{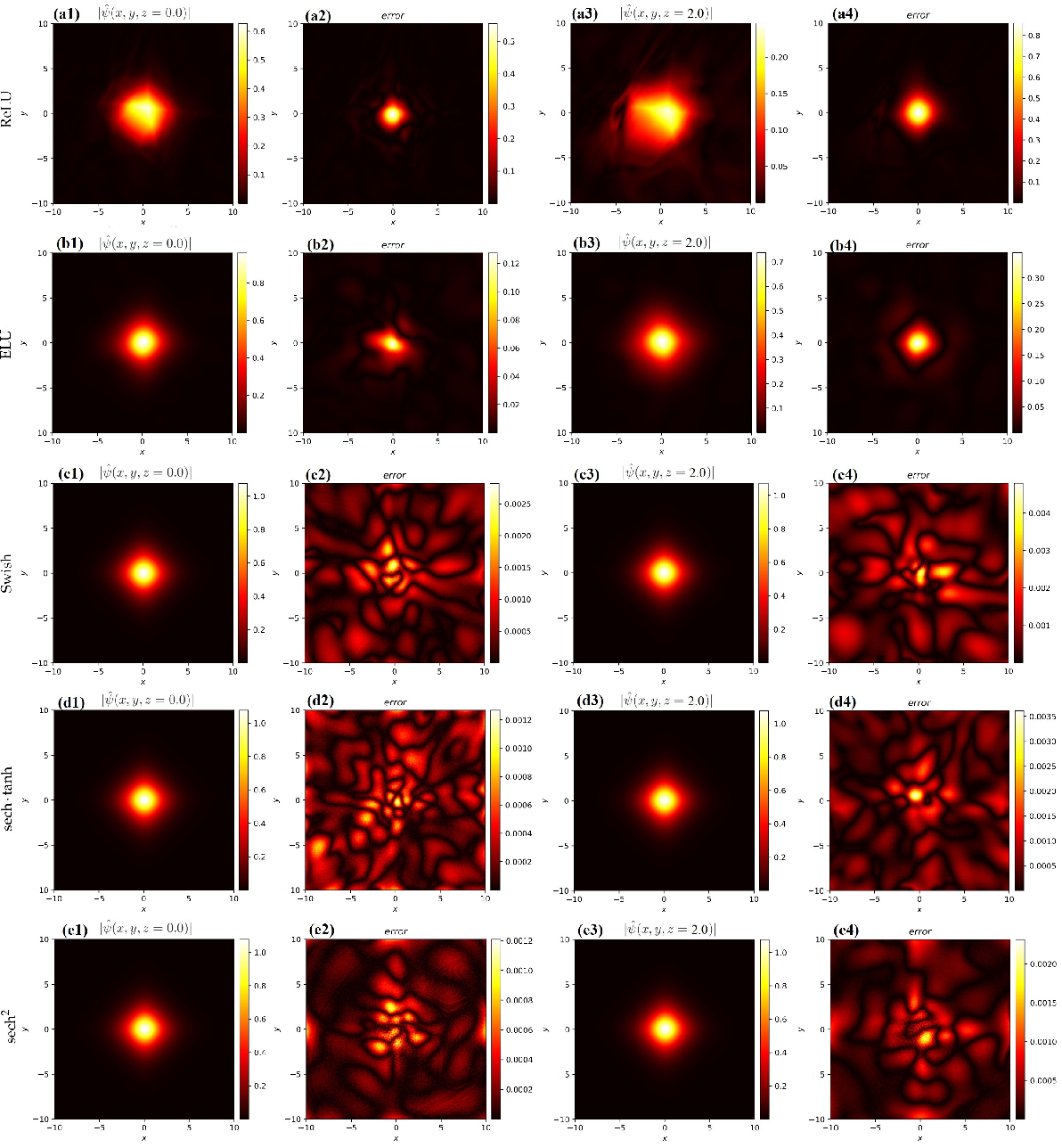}}}\hspace{-0.35in}
\vspace{0.1in}
\caption{\small The results for 2D Scarf-II potential with different activation functions. Rows one to five, respectively, represent the activation functions ReLU, ELU, Swish, sech$\cdot$tanh,
and sech$^2$. Columns one to four, respectively, represent the predicted magnitudes of solutions,  error values in magnitudes of solutions, and solitons at different propagation distances $z = 0$ and $2$.}
  \label{scarf2D_act}
\end{figure*}

The $\mathbb{L}^2$-norm errors in approximating $p(\bx,z)$, $q(\bx,z)$ and $\psi(\bx,z)$ for all cases and also for those activation functions are exhibited in Table~\ref{table3_1}. For example, for the data-driven solitons of the SNLSE with $\PT$ Scarf-II potential, we consider those general activation functions to investigate the effect of other activation functions in learning the $\PT$-symmetric soliton solutions. Especially, considering that the Scarf-II potential consists of hyperbolic secant and hyperbolic tangent functions, we use a new class of activation functions, i.e., sech$\cdot$tanh and ${\rm sech}^2$, to see if they can work better. We choose the same network structure and training steps as in Sec. \ref{sec2} (similarly hereinafter).
The predicted values and the error values in the case of SNLSE with $\PT$ Scarf-II potential in approximating the soliton solutions with five different activation functions are exhibited in Figs.~\ref{scarf1D_act} and \ref{scarf2D_act}.

\begin{figure*}[!t]
    \centering
\vspace{-0.15in}
  {\scalebox{0.75}[0.7]{\includegraphics{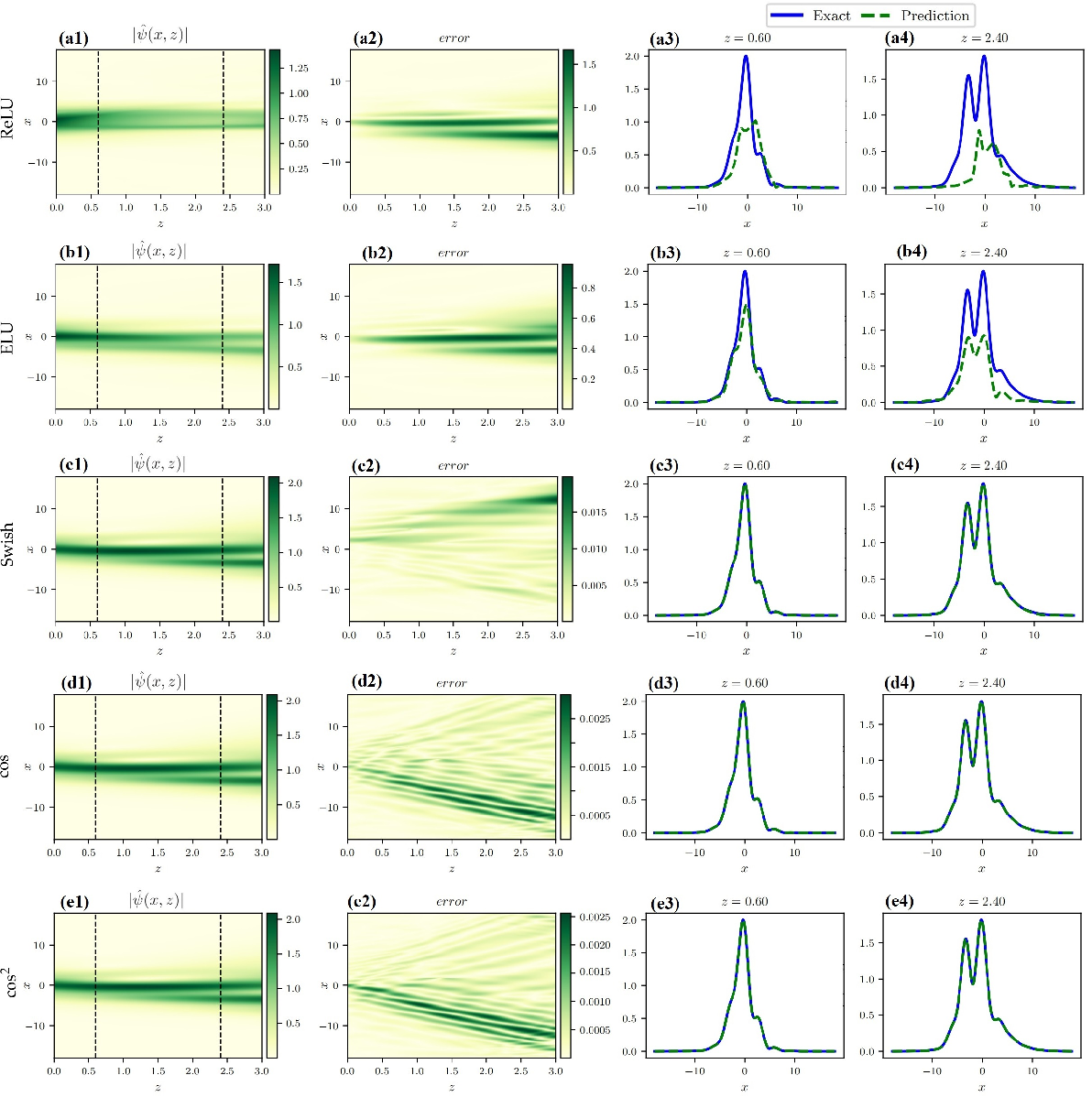}}}\hspace{-0.35in}
\vspace{0.1in}
\caption{\small The results for 1D $\PT$ period potential with different activation functions. Rows one to five, respectively, represent the activation functions ReLU, ELU, Swish, cos, and cos$^2$. Columns one to four, respectively, represent the predicted magnitudes of solutions, error values in magnitudes of solutions, and solitons at different propagation distances $z = 0.6$ and $2.4$.}
  \label{period1D_act}
\end{figure*}

\begin{figure*}[!t]
    \centering
\vspace{-0.15in}
  {\scalebox{0.75}[0.7]{\includegraphics{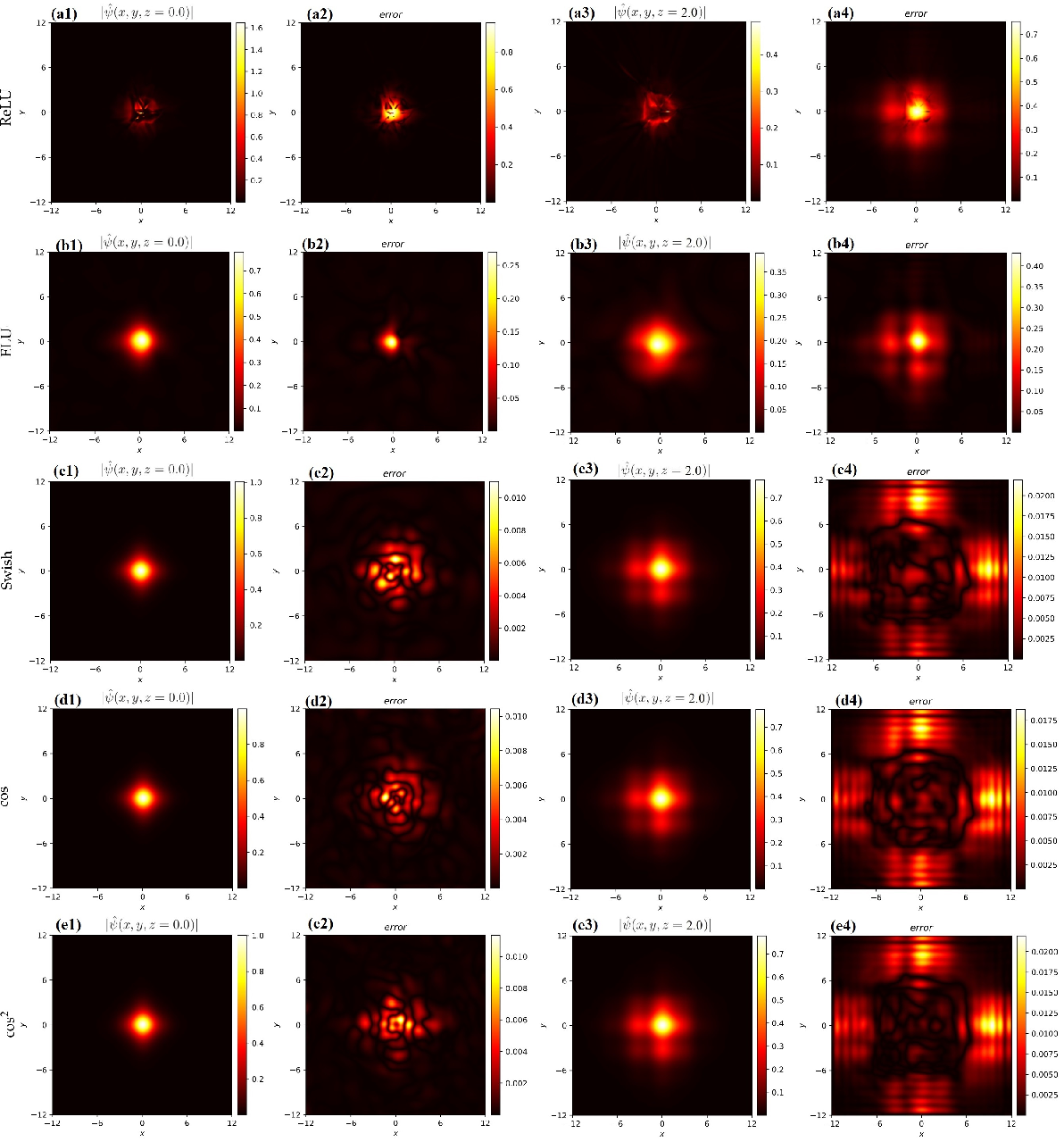}}}\hspace{-0.35in}
\vspace{0.1in}
\caption{\small The results for 2D $\PT$ period potential with different activation functions. Rows one to five, respectively, represent the activation functions ReLU, ELU, Swish, cos, and cos$^2$. Columns one to four, respectively, represent the predicted magnitudes of solutions, error values in magnitudes of solutions, and solitons at different propagation distances $z = 0$ and $2$.}
  \label{period2D_act}
\end{figure*}

Since ReLU is an unsmooth function, thus $\mathbb{L}^2$-norm errors with ReLU as the activation function are very high in predicting soliton solution compared with other activation functions, which are shown in Figs.~\ref{scarf1D_act}(a1-a4) and Figs.~\ref{scarf2D_act}(a1-a4). Moreover, since ELU is not second-order continuous, and the second derivative for the SNLSE is required. Therefore $\mathbb{L}^2$-norm errors with ELU as the activation function are also very high (see Figs.~\ref{scarf1D_act}(b1-b4) and Figs.~\ref{scarf2D_act}(b1-b4)).
And for the activation functions, cos and ${\rm cos}^2$, their $\mathbb{L}^2$-norm errors are of the order of $10^{-4}$ for the 1D Scarf-II potential, while $\mathbb{L}^2$ norm errors are of the order of $10^{-2}$ for the 2D Scarf-II potential.
However, when the activation functions are the hyperbolic functions, $\mathbb{L}^2$-norm errors are lower and of the order of $10^{-4}$ for the 1D Scarf-II potential and $10^{-3}$ for the 2D Scarf-II potential. In particular, it is obviously that the error value is slightly lower for the cases of sech$\cdot$tanh and ${\rm sech}^2$ (see Figs.~\ref{scarf1D_act}(d1-d4, e1-e4) and Figs.~\ref{scarf2D_act}(d1-d4, e1-e4)). The reason for this phenomenon is not only that the $\PT$ Scarf-II potential is composed of hyperbolic functions (sech and tanh), but also that the real and imaginary parts are quadratic with respect to the hyperbolic function. The results for other activation functions can be found in Table \ref{table3_1}.

For the data-driven solutions of the SNLSE with $\PT$ period potential, we consider the same activation functions to study their effects in learning the non-stationary solutions. Similarly, from Figs.~\ref{period1D_act}(a1-a4, b1-b4) and Figs.~\ref{period2D_act}(a1-a4, b1-b4), we can see that the approximations with ReLU and ELU activation functions do not fit well with the original results. But when the activation functions are taken as periodic functions (e.g., cos and $\cos^2$), the predicted results become better by comparing them with the predictions learned by other activation functions since the squared error values come out less. Especially, for the activation function $\cos^2$, $\mathbb{L}^2$-norm errors are lower and of
the order of $10^{-3}$. The results for other activation functions can be found in Table \ref{table3_1}.

Therefore, we may conclude that selecting the activation function according to the structures of solutions and equations usually can achieve the better effect. Furthermore we also find that choosing the Swish or rational functions ($F_5$ and $F_6$) as the activation function also usually achieves better results, which can provide one with more options for activation functions (see Table \ref{table3_1} and Figs.~\ref{scarf1D_act}, \ref{scarf2D_act}, \ref{period1D_act} and \ref{period2D_act} for more details).

\subsection{Influence of structure of the neural networks}

As we know, the parameters in the neural networks have great influence on the network performance, such as learning rate, training step, width of the network (number of hidden layers), and depth of the network (number of neurons in each layer). Here we mainly consider the effects of the latter two, that is, the impacts of number of hidden layers and neurons in each layer for all four potentials with tanh as an activation function, and the outcomes are exhibited in Tables \ref{table3_2} and \ref{table3_3}. In Table \ref{table3_2}, the number of neurons in each layer is fixed at 100. And the sample points and training steps are same as in Sec. \ref{sec2}. It is obviously that the network performance with single hidden layer is very poor in any case. And with the increase of the number of hidden layers, the corresponding network performance improves in any case, i.e., $\mathbb{L}^2$ norm errors decrease. Therefore, in this paper, we fix the number of hidden layers at 4 to achieve the better network performance.

On the other hand, Table \ref{table3_3} shows the impact of different number of neurons in the hidden layers on the network performance, where the activation function is tanh, and four hidden layers are taken. We can see that the network performance with ten neurons is very poor in four cases, i.e., the $\mathbb{L}^2$ norm errors are very high. And with the increase of number of neurons, the error values are decreasing. However for some cases they will get to a certain precision and oscillate back and forth since more neurons increases the size of the weight and bias matrices and the model needs to optimize the more number of parameters.

\begin{table}[!t]
\begin{center}\small
\caption{\small $\mathbb{L}^2$-norm errors of $p(\bx,z)$, $q(\bx,z)$ and $\psi(\bx,z)$ of the SNLSE with 1D/2D $\PT$ Scarf-II or periodic potential for different number of hidden layers.}\label{table3_2}
\vspace{0.02in}
\begin{tabular}{ cccccc }
\hline
\hline\\[-2ex]
 &  &  \multicolumn{4}{c}{Number of hidden layers}  \\[1ex]
\cline{3-6}\\[-2ex]
    $\PT$-symmtric potentials    & $\mathbb{L}^2$-norm errors   & 1   & 2  & 3 & 4  \\[0.5ex]
\hline

\rule{0pt}{13pt}                &$p$     & $6.508 \cdot 10^{-2}$    & $2.612 \cdot 10^{-3}$  & $7.683 \cdot 10^{-4}$& $5.576 \cdot 10^{-4}$\\
\rule{0pt}{13pt}  1D Scarf-II   &$q$     & $6.185 \cdot 10^{-2}$    & $2.621 \cdot 10^{-3}$  & $6.660 \cdot 10^{-4}$& $5.294 \cdot 10^{-4}$\\
\rule{0pt}{13pt}               &$\psi$   & $4.062 \cdot 10^{-2}$    & $1.726 \cdot 10^{-3}$  & $4.961 \cdot 10^{-4}$& $4.019 \cdot 10^{-4}$\\[0.5ex]
\hline

\rule{0pt}{13pt}                &$p$     & $1.630 \cdot 10^{-1}$    & $3.523 \cdot 10^{-2}$  & $1.224 \cdot 10^{-2}$& $7.593 \cdot 10^{-3}$\\
\rule{0pt}{13pt}  2D Scarf-II   &$q$     & $1.764 \cdot 10^{-1}$    & $3.983 \cdot 10^{-2}$  & $1.241 \cdot 10^{-2}$& $8.136 \cdot 10^{-3}$\\
\rule{0pt}{13pt}               &$\psi$   & $1.405 \cdot 10^{-1}$    & $3.112 \cdot 10^{-2}$  & $9.320 \cdot 10^{-3}$& $6.131 \cdot 10^{-3}$\\[0.5ex]
\hline

\rule{0pt}{13pt}                &$p$     & $4.523 \cdot 10^{-2}$    & $2.136 \cdot 10^{-2}$  & $1.171 \cdot 10^{-2}$& $7.719 \cdot 10^{-3}$\\
\rule{0pt}{13pt}   1D Periodic  &$q$     & $5.612 \cdot 10^{-2}$    & $2.377 \cdot 10^{-2}$  & $9.982 \cdot 10^{-3}$& $6.739 \cdot 10^{-3}$\\
\rule{0pt}{13pt}               &$\psi$   & $4.194 \cdot 10^{-2}$    & $1.821 \cdot 10^{-2}$  & $8.020 \cdot 10^{-3}$& $5.400 \cdot 10^{-3}$\\[0.5ex]
\hline

\rule{0pt}{13pt}                &$p$     & $2.861 \cdot 10^{-1}$    & $3.265 \cdot 10^{-1}$  & $8.923 \cdot 10^{-2}$& $5.697 \cdot 10^{-2}$\\
\rule{0pt}{13pt}   2D Periodic  &$q$     & $3.115 \cdot 10^{-1}$    & $1.809 \cdot 10^{-1}$  & $7.778 \cdot 10^{-2}$& $5.703 \cdot 10^{-2}$\\
\rule{0pt}{13pt}               &$\psi$   & $2.476 \cdot 10^{-1}$    & $1.597 \cdot 10^{-1}$  & $6.589 \cdot 10^{-2}$& $4.538 \cdot 10^{-2}$\\[0.5ex]
\hline
\hline
\end{tabular}
\end{center}
\end{table}

\begin{table}[!t]
\begin{center}\small
\caption{\small$\mathbb{L}^2$-norm errors of $p(\bx,z)$, $q(\bx,z)$ and $\psi(\bx,z)$ of the SNLSE with 1D/2D $\PT$ Scarf-II or periodic potential for different number of neurons in each hidden layer.}\label{table3_3}
\vspace{0.02in}
\begin{tabular}{ cccccc }
\hline
\hline\\[-2ex]
 &  &  \multicolumn{4}{c}{Number of neurons in each hidden layer}  \\[1ex]
\cline{3-6}\\[-2ex]
    $\PT$-symmetric potentials    & $\mathbb{L}^2$-norm errors \quad  & 10   & 40  & 70 & 100  \\[0.5ex]
\hline

\rule{0pt}{13pt}                &$p$     & $7.912 \cdot 10^{-1}$    & $6.059 \cdot 10^{-4}$  & $3.371 \cdot 10^{-4}$& $5.576 \cdot 10^{-4}$\\
\rule{0pt}{13pt}  1D Scarf-II   &$q$     & $8.752 \cdot 10^{-1}$    & $5.389 \cdot 10^{-4}$  & $3.551 \cdot 10^{-4}$& $5.294 \cdot 10^{-4}$\\
\rule{0pt}{13pt}               &$\psi$   & $6.895 \cdot 10^{-1}$    & $4.027 \cdot 10^{-4}$  & $2.633 \cdot 10^{-4}$& $4.019 \cdot 10^{-4}$\\[0.5ex]
\hline

\rule{0pt}{13pt}                &$p$     & $5.799 \cdot 10^{-1}$    & $2.938 \cdot 10^{-2}$  & $3.345 \cdot 10^{-2}$& $7.593 \cdot 10^{-3}$\\
\rule{0pt}{13pt}  2D Scarf-II   &$q$     & $6.756 \cdot 10^{-1}$    & $2.974 \cdot 10^{-2}$  & $3.907 \cdot 10^{-2}$& $8.136 \cdot 10^{-3}$\\
\rule{0pt}{13pt}               &$\psi$   & $6.058 \cdot 10^{-1}$    & $2.279 \cdot 10^{-2}$  & $3.001 \cdot 10^{-2}$& $6.131 \cdot 10^{-3}$\\[0.5ex]
\hline

\rule{0pt}{13pt}                &$p$     & $8.623 \cdot 10^{-1}$    & $1.891 \cdot 10^{-3}$  & $1.768 \cdot 10^{-3}$& $7.719 \cdot 10^{-3}$\\
\rule{0pt}{13pt}   1D Periodic  &$q$     & $7.346 \cdot 10^{-1}$    & $1.730 \cdot 10^{-3}$  & $1.625 \cdot 10^{-3}$& $6.739 \cdot 10^{-3}$\\
\rule{0pt}{13pt}               &$\psi$   & $7.057 \cdot 10^{-1}$    & $1.256 \cdot 10^{-3}$  & $1.083 \cdot 10^{-3}$& $5.400 \cdot 10^{-3}$\\[0.5ex]
\hline

\rule{0pt}{13pt}                &$p$     & $1.306 \cdot 10^{-1}$    & $5.280 \cdot 10^{-2}$  & $4.577 \cdot 10^{-2}$& $5.697 \cdot 10^{-2}$\\
\rule{0pt}{13pt}   2D Periodic  &$q$     & $1.385 \cdot 10^{-1}$    & $5.480 \cdot 10^{-2}$  & $4.601 \cdot 10^{-2}$& $5.703 \cdot 10^{-2}$\\
\rule{0pt}{13pt}               &$\psi$   & $1.129 \cdot 10^{-1}$    & $4.168 \cdot 10^{-2}$  & $3.589 \cdot 10^{-2}$& $4.538 \cdot 10^{-2}$\\[0.5ex]
\hline
\hline
\end{tabular}
\end{center}
\end{table}

\begin{table}[!t]
\begin{center}\small
\caption{\small $\mathbb{L}^2$-norm errors of $p(\bx,z)$, $q(\bx,z)$ and $\psi(\bx,z)$ of the SNLSE with 1D/2D $\PT$ Scarf-II or periodic potential for different number of collocation points.}\label{table3_4}
\vspace{0.02in}
\begin{tabular}{ cccccc }
\hline
\hline\\[-2ex]
 &  &  \multicolumn{4}{c}{Number of collocation points}  \\[0.2ex]
\cline{3-6}\\[-2ex]
    $\PT$-symmetric potentials    &$\mathbb{L}^2$-norm errors   & 2000  & 5000  & 10000 & 20000  \\[1ex]
\hline

\rule{0pt}{13pt}                &$p$     & $5.576 \cdot 10^{-4}$    & $4.571 \cdot 10^{-4}$  & $5.464 \cdot 10^{-4}$& $4.725 \cdot 10^{-4}$\\
\rule{0pt}{13pt}  1D Scarf-II   &$q$     & $5.294 \cdot 10^{-4}$    & $4.894 \cdot 10^{-4}$  & $5.010 \cdot 10^{-4}$& $4.462 \cdot 10^{-4}$\\
\rule{0pt}{13pt}               &$\psi$   & $4.019 \cdot 10^{-4}$    & $3.355 \cdot 10^{-4}$  & $3.713 \cdot 10^{-4}$& $3.396 \cdot 10^{-4}$\\[0.5ex]
\hline

\rule{0pt}{13pt}                &$p$     & $1.430 \cdot 10^{-2}$    & $9.857 \cdot 10^{-3}$  & $7.849 \cdot 10^{-3}$& $9.745 \cdot 10^{-3}$\\
\rule{0pt}{13pt}  2D Scarf-II   &$q$     & $1.500 \cdot 10^{-2}$    & $1.041 \cdot 10^{-2}$  & $7.656 \cdot 10^{-3}$& $1.029 \cdot 10^{-2}$\\
\rule{0pt}{13pt}               &$\psi$   & $1.277 \cdot 10^{-2}$    & $7.941 \cdot 10^{-3}$  & $5.756 \cdot 10^{-3}$& $7.963 \cdot 10^{-3}$\\[0.5ex]
\hline

\rule{0pt}{13pt}                &$p$     & $9.630 \cdot 10^{-2}$    & $8.757 \cdot 10^{-3}$  & $8.740 \cdot 10^{-3}$& $8.145 \cdot 10^{-3}$\\
\rule{0pt}{13pt}   1D Periodic  &$q$     & $7.758 \cdot 10^{-2}$    & $6.802 \cdot 10^{-3}$  & $7.057 \cdot 10^{-3}$& $6.425 \cdot 10^{-3}$\\
\rule{0pt}{13pt}               &$\psi$   & $6.654 \cdot 10^{-2}$    & $5.911 \cdot 10^{-3}$  & $5.953 \cdot 10^{-3}$& $5.392 \cdot 10^{-3}$\\[0.5ex]
\hline

\rule{0pt}{13pt}                &$p$     & $1.111 \cdot 10^{0}$     & $3.706 \cdot 10^{-1}$  & $9.982 \cdot 10^{-2}$& $7.297 \cdot 10^{-2}$\\
\rule{0pt}{13pt}   2D Periodic  &$q$     & $7.744 \cdot 10^{-1}$    & $3.147 \cdot 10^{-1}$  & $9.024 \cdot 10^{-2}$& $7.565 \cdot 10^{-2}$\\
\rule{0pt}{13pt}               &$\psi$   & $3.217 \cdot 10^{-1}$    & $1.762 \cdot 10^{-1}$  & $7.680 \cdot 10^{-2}$& $5.598 \cdot 10^{-2}$\\[0.5ex]

\hline
\hline
\end{tabular}
\end{center}
\end{table}

\subsection{Influence of number of sampling points}

Finally, we discuss the influence of number  of sampling points on the neural network performance. On the one hand, with respect to different number of collocation points, we randomly choose different points in the domain $\Omega \times [0, Z]$ with tanh as an activation function for four different kinds of $\PT$-symmetric potentials, where the numbers of initial and boundary points, training steps and network structure are same as ones in Sec. \ref{sec2}.  According to the results shown in Table \ref{table3_4}, we can see that when more collocation points are taken inside the domain $\Omega \times [0, Z]$,
the model can be trained with more points to lead to high accurate solutions. However, when the number of collocation points reaches a certain value, the error does not decrease significantly but fluctuates within a certain range. Therefore we need to select a suitable number of collocation points to save training costs and achieve higher accuracy. From Table \ref{table3_4}, 2000 collocation points is enough for the 1D Scarf-II potential to achieve high accuracy, while for the 2D Periodic potential more points are needed. In general, higher dimensional cases require more collocation points.

On the other hand, by changing the number of initial and boundary sample points, the $\mathbb{L}^2$ norm errors of $p(\bx,z)$, $q(\bx,z)$ and $\psi(\bx,z)$ of the SNLSE with $\PT$ Scarf-II  and periodic potentials in 1D and 2D cases are exhibited in Table \ref{table3_5}. The quantity ratio of initial sample points to boundary sample points is $2:1$. For example, the number 30 in the table represents that there are 20 points on the initial data and 10 points on boundary data. And other parameters including the number of collocation points and training steps, network structure as well as activation function are same as in Sec. \ref{sec2}. As is shown in Table \ref{table3_5}, the $\mathbb{L}^2$ norm errors will be lower when increasing the number of points in most instances, while in some cases, the error values vary between low and high values. For example, for the 1D Scarf-II potential, the number of initial and boundary points has little effect on the errors being of the order of $10^{-4}$. However,
in the case of high dimension, the error values are very high with low number of initial and boundary points. Therefore,
the higher-dimensional case also requires more initial and boundary sampling points in general.

\begin{table}[!t]
\begin{center}
\caption{\small $\mathbb{L}^2$-norm errors of $p(\bx,z)$, $q(\bx,z)$ and $\psi(\bx,z)$ of the NLSE with SN and 1D/2D $\PT$ Scarf-II or periodic potential for different number of initial and boundary points.}\label{table3_5}
\vspace{0.02in}
{\small\begin{tabular}{ ccccccc }
\hline
\hline\\[-2ex]
 &  &  \multicolumn{5}{c}{Number of initial and boundary points}  \\[0.2ex]
\cline{3-7}\\[-2ex]
    $\PT$-symmetric potentials    & $\mathbb{L}^2$-norm errors   & 30  & 90  & 150 & 300 & 450 \\[0.2ex]
\hline

\rule{0pt}{13pt}                &$p$     & $8.323 \cdot 10^{-4}$    & $4.898 \cdot 10^{-4}$  & $5.099\cdot 10^{-4}$& $3.696 \cdot 10^{-4}$ & $4.763 \cdot 10^{-4}$\\
\rule{0pt}{13pt}  1D Scarf-II   &$q$     & $8.223 \cdot 10^{-4}$    & $5.003 \cdot 10^{-4}$  & $5.865\cdot 10^{-4}$& $4.151 \cdot 10^{-4}$ & $3.512 \cdot 10^{-4}$\\
\rule{0pt}{13pt}               &$\psi$   & $5.629 \cdot 10^{-4}$    & $3.432 \cdot 10^{-4}$  & $3.867\cdot 10^{-4}$& $2.774 \cdot 10^{-4}$ & $2.836 \cdot 10^{-4}$\\[0.5ex]
\hline

\rule{0pt}{13pt}                &$p$     & $8.167 \cdot 10^{-1}$    & $2.252 \cdot 10^{-2}$  & $2.055 \cdot 10^{-2}$& $3.608 \cdot 10^{-2}$ & $1.419 \cdot 10^{-2}$\\
\rule{0pt}{13pt}  2D Scarf-II   &$q$     & $7.697 \cdot 10^{-1}$    & $2.047 \cdot 10^{-2}$  & $1.803 \cdot 10^{-2}$& $3.052 \cdot 10^{-2}$ & $1.395 \cdot 10^{-2}$\\
\rule{0pt}{13pt}               &$\psi$   & $6.944 \cdot 10^{-1}$    & $1.735 \cdot 10^{-2}$  & $1.601 \cdot 10^{-2}$& $2.644 \cdot 10^{-2}$ & $1.514 \cdot 10^{-2}$\\[0.5ex]
\hline

\rule{0pt}{13pt}                &$p$     & $2.733 \cdot 10^{-1}$    & $1.296 \cdot 10^{-2}$  & $9.235 \cdot 10^{-3}$& $7.367 \cdot 10^{-3}$ & $7.548 \cdot 10^{-3}$\\
\rule{0pt}{13pt}   1D Periodic  &$q$     & $2.210 \cdot 10^{-1}$    & $1.137 \cdot 10^{-2}$  & $7.610 \cdot 10^{-3}$& $6.069 \cdot 10^{-3}$ & $6.182 \cdot 10^{-3}$\\
\rule{0pt}{13pt}               &$\psi$   & $1.731 \cdot 10^{-1}$    & $9.255 \cdot 10^{-3}$  & $6.667 \cdot 10^{-3}$& $4.995 \cdot 10^{-3}$ & $5.260 \cdot 10^{-3}$\\[0.5ex]
\hline

\rule{0pt}{13pt}                &$p$     & $9.924 \cdot 10^{-1}$    & $3.506 \cdot 10^{-1}$  & $1.277 \cdot 10^{-1}$& $9.353 \cdot 10^{-2}$ & $9.369 \cdot 10^{-2}$\\
\rule{0pt}{13pt}   2D Periodic  &$q$     & $9.599 \cdot 10^{-1}$    & $3.004 \cdot 10^{-1}$  & $1.284 \cdot 10^{-1}$& $9.973 \cdot 10^{-2}$ & $9.834 \cdot 10^{-2}$\\
\rule{0pt}{13pt}               &$\psi$   & $9.582 \cdot 10^{-1}$    & $2.788 \cdot 10^{-1}$  & $9.662 \cdot 10^{-2}$& $6.686 \cdot 10^{-2}$ & $6.718 \cdot 10^{-2}$\\[0.5ex]

\hline
\hline
\end{tabular}}
\end{center}
\end{table}

\section{Data-driven $\PT$ potentials discovery in the SNLSE}\label{sec3}

 In this section, we consider the inverse problem of the SNLSE about the data-driven $\PT$-symmetric potential discovery. Since the classical PINNs can only be used to learn the parameters of equations~\cite{pinn}, thus we here propose a
modified PINNs (mPINNs) method based on the PINNs to identify the whole $\PT$ potential $V(\bx)+iW(\bx)$ of the 1D and 2D NNLSEs rather than just potential parameters. In particular, for the stationary nonlinear modes, the physical information part of the network can be improved to reduce training time and sample set. In Sec. 4.1, we present the mPINNs framework for the general solutions, and the PINNs framework for $\PT$ potential discovery for the stationary solutions. In Sec. 4.2, we use two frameworks, namely PINNs and mPINNs, to powerfully learn the 1D and 2D $\PT$ Scarf-II potentials from the stationary solutions and non-stationary solutions. Moreover, we also find that the mPINNs can be used to study the inverse problem about $\PT$ Scarf-II potential dependent on propagation distance $z$ in the 1D SNLSE. Similarly, in Sec. 4.3, we use two frameworks, namely PINNs and mPINNs, to powerfully learn the 1D and 2D  $\PT$ periodic potentials from the stationary solutions and non-stationary solutions. Moreover, we also find that the mPINNs can be used to study the inverse problem about $\PT$ periodic potential dependent on propagation distance $z$ in the 1D SNLSE.

\begin{figure*}[!t]
    \centering
  {\scalebox{0.8}[0.82]{\includegraphics{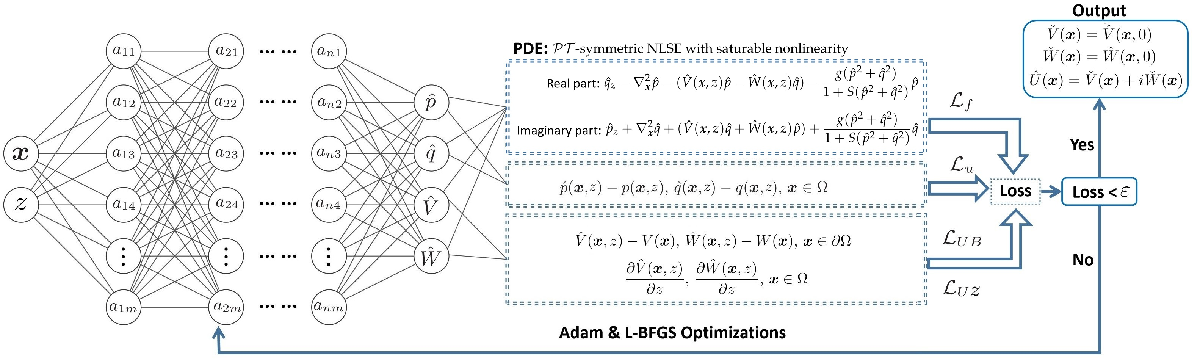}}}\hspace{-0.35in}
\vspace{0.15in}
\caption{\small The mPINNs deep learning framework for inverse problem of the $\PT$-symmetric potential discovery.}
  \label{net2}
\end{figure*}

\subsection{The mPINNs framework for inverse problems}

\subsubsection{The mPINNs framework for the general solution}

 For the inverse problem of the $\PT$-symmetric SNLSE (\ref{nls}), the complex potential $U(\bx)=V(\bx)+iW(\bx)$ is unknown function to be determined. Similarly, let $\psi(\bx,z)=p(\bx,z)+iq(\bx,z)$, where $p(\bx,z)$ and $q(\bx,z)$ being its real and imaginary parts, respectively. We still use a fully connected neural network and take $\left[\hat{p}(\bx,z),\,\hat{q}(\bx,z),\,\hat{V},\,\hat{W}\right]$ as the outputs of the network. Since $\bx,z$ are taken as the inputs of the network, then $\hat{V}=\hat{V}(\bx,z)$ and $\hat{W}=\hat{W}(\bx,z)$. However the $\PT$-symmetric potential we discussed in Secs. 2 and 3 is independent of the propagation distance $z$. Then we need to add $\frac{\partial \hat{V}(\bx,z)}{\partial z}$ and $\frac{\partial \hat{W}(\bx,z)}{\partial z}$ to the loss function.
Therefore, we use a complex-valued deep neural network to approximate $\psi(\bx,z)$ and $U(\bx)=V(\bx)+iW(\bx)$, and then the mPINN $m\mathcal{F}(\bx, z)$ is given by
\bee\label{mF}
m\mathcal{F}(\bx, z):=-m\mathcal{F}_p(\bx, z)+im\mathcal{F}_q(\bx, z)= i\hat{\psi}_z+\nabla_{\bx}^2\hat{\psi}+[\hat{V}(\bx,z)+i\hat{W}(\bx,z)]\hat{\psi}+\frac{g|\hat{\psi}|^2}{1+S|\hat{\psi}|^2}\hat{\psi},
\ene
with $-m\mathcal{F}_p(\bx, z)$ and $m\mathcal{F}_q(\bx, z)$ being its real and imaginary parts, respectively, written as
  \begin{equation}\label{mF2}
 \begin{array}{l}
     \displaystyle
     \displaystyle m\mathcal{F}_p(\bx, z):= \hat{q}_z-\nabla^2_{\bx}\hat{p}-\hat{V}(\bx,z)\hat{p}+\hat{W}(\bx,z)\hat{q}-\frac{g(\hat{p}^2+\hat{q}^2)}{1+S(\hat{p}^2+\hat{q}^2)}\hat{p},\v\\
     \displaystyle m\mathcal{F}_q(\bx, z):= \hat{p}_z+\nabla^2_{\bx}\hat{q}+\hat{V}(\bx,z)\hat{q}+\hat{W}(\bx,z)\hat{p}+\frac{g(\hat{p}^2+\hat{q}^2)}{1+S(\hat{p}^2+\hat{q}^2)}\hat{q}.
 \end{array}
\end{equation}
And they can be trained by minimizing the total mean squared error loss containing four parts
\begin{equation}\label{mloss}
  m\mathcal{T\!L}=m\mathcal{L}_f+m\mathcal{L}_U+m\mathcal{L}_{UB}+m\mathcal{L}_{UZ},
\end{equation}
with
\begin{equation}\label{mlfib}
 \begin{array}{l}
     \displaystyle\quad m\mathcal{L}_f=\frac{1}{N_f}\sum_{\ell=1}^{N_f}\left(|m\mathcal{F}_p(\bx_f^\ell,z_f^\ell)|^2
     +|m\mathcal{F}_q(\bx_f^\ell, z_f^\ell)|^2\right),\v\\
     \displaystyle\quad m\mathcal{L}_U=\frac{1}{N_f}\sum_{\ell=1}^{N_f}\left(|\hat{p}(\bx_f^\ell, z_f^\ell)-p^\ell|^2+|\hat{q}(\bx_f^\ell, z_f^\ell)-q^\ell|^2\right),\v\\
     \displaystyle\quad m\mathcal{L}_{UB}=\frac{1}{N_B}\sum_{\ell=1}^{N_B}\left(|\hat{V}(\bx_B^\ell,z_B^\ell)-V(\bx_B^\ell)|^2
     +|\hat{W}(\bx_B^\ell,z_B^\ell)-W(\bx_B^\ell)|^2\right),\v\\
     \displaystyle\quad m\mathcal{L}_{UZ}=\frac{1}{N_f}\sum_{\ell=1}^{N_f}\left(\left|\frac{\partial \hat{V}(\bx_f^\ell,z_f^\ell)}{\partial z}\right|^2+\left|\frac{\partial \hat{W}(\bx_f^\ell,z_f^\ell)}{\partial z}\right|^2\right),
 \end{array}
\end{equation}
where $\{\bx_f^\ell,z_f^\ell,p^\ell, q^\ell\}_\ell^{N_f}$ are connected with the training data on the real part and imaginary part of exact solution $p(\bx,z), q(\bx,z)$  with $\psi(\bx_f^\ell,z_f^\ell)=p^\ell+iq^\ell$ in domain $\Omega\times[0,Z]$, and $\{\bx_B^\ell,z_B^\ell,V(\bx_B^\ell),W(\bx_B^\ell)\}_\ell^{N_B}$ are linked with the randomly selected boundary training data of potential $U(\bx)=V(\bx)+iW(\bx)$ in domain $\partial\Omega\times[0,Z]$. Finally, we set $\check{V}(\bx)=\hat{V}(\bx,0)$, $\check{W}(\bx)=\hat{W}(\bx,0)$ and take $\hat{U}(\bx)=\check{V}(\bx)+i\check{W}(\bx)$ as the predicted potential (see Fig.~\ref{net2} for the mPINNs scheme in detail). It should be mentioned that for the given boundary conditions, the potential function is uniquely defined via mPINNS. but this situation will provide one with more choices about the potential for the experimental researchers. The main steps of the mPINNs method determining the $\PT$-symmetric potentials of SNLSE (\ref{nls}) are
given in Table~\ref{table-m}.

\begin{table}[h]
\centering
\caption{The  mPINNs method determining the $\PT$-symmetric potentials of SNLSE (\ref{nls}).}
\begin{tabularx}{\textwidth}{lX}
\hline\hline
Step &  Instruction \\
\hline
1 & Establishing a fully-connected neural network NN$(\bx, z; W, B)$ with initialized parameters $W = \{w_j\}_{1}^{n+1}$ and $B = \{b_j\}_1^{n+1}$ being the weights and bias, and the mPINNs $m\mathcal{F}(\bx, z)$ is given by Eq.~(\ref{mF}); \\
2 & Constructing the two training data sets in domain $\Omega\times[0,Z]$ and boundary $\partial\Omega\times[0,Z]$
for the boundary value and considered model; \\
3& Constructing a training loss function $m\mathcal{T\!L}$ given by Eq.~(\ref{mloss}) by summing the MSE containing four parts $m\mathcal{L}_f$, $m\mathcal{L}_U$, $m\mathcal{L}_{UB}$ and $m\mathcal{L}_{UZ}$; \\
4& Train the NN to optimize the parameters $\{W, B\}$ by minimizing the loss function $m\mathcal{TL}$ in terms of
the Adam \& L-BFGS optimization algorithm.  \\
\hline\hline
\end{tabularx}
\label{table-m}
\end{table}

\begin{figure*}[!t]
    \centering
  {\scalebox{0.8}[0.82]{\includegraphics{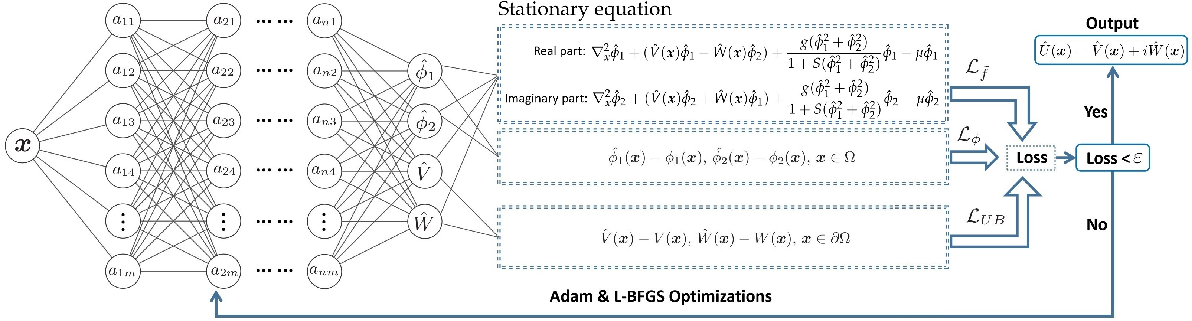}}}\hspace{-0.35in}
\vspace{0.15in}
\caption{\small The PINNs scheme for inverse problem of $\PT$-symmetric potential discovery related to the stationary equation.}
  \label{net3}
\end{figure*}

\subsubsection{The PINNs framework for $\PT$ potential discovery for the stationary solution}

For the inverse problems of PDEs via the PINNs, the data-driven parameter discovery, not function discovery, is usually studied. Here we extend the PINNs to study the varying coefficients (i.e., complex potential $V(x)+iW(x)$), not some parameters, of Eq.~(\ref{nls}) by considering its stationary equation (i.e., via $\psi(\bx,z)=\phi(\bx)e^{i\mu z}$ with $\phi(\bx)\in\mathbb{C}[\bx]$)
\bee\label{ode}
  \nabla_{\bx}^2\phi+[V(\bx)+iW(\bx)]\phi+\frac{g|\phi|^{2}}{1+S|\phi|^{2}}\phi=\mu \phi.
\ene
{\bf Remark.} {\it If one PDE can not reduce to the corresponding stationary equation (\ref{ode}), then one can not directly use the PINNs to study the varying coefficients with spatial variable. In this case, one can use the above-mentioned mPINNs.}



\v Let $\phi(\bx)=\phi_1(\bx)+i\phi_2(\bx)$ with $\phi_1(\bx),\,\phi_2(\bx)\in\mathbb{R}[\bx]$. Then, similarly, we still use a fully-connected network and take $[\hat{\phi}_1(\bx)$,\, $\hat{\phi}_2(\bx),$ \, $\hat{V}$, \, $\hat{W}]$ as the outputs of the network. Then the PINNs related to the physical information of the corresponding stationary equation $\mathcal{F}(\bx)$ is
\bee\label{FF}
\mathcal{F}(\bx):=\mathcal{F}_1(\bx)+i\mathcal{F}_2(\bx)= \nabla_{\bx}^2\hat{\phi}+[\hat{V}(\bx)+i\hat{W}(\bx)]\hat{\phi}+\frac{g|\hat{\phi}|^{2}}{1+S|\hat{\phi}|^{2}}\hat{\phi}-\mu \hat{\phi}
     \ene
 with $\mathcal{F}_1(\bx)$ and $\mathcal{F}_2(\bx)$ being its real and imaginary parts, written as
 \begin{equation}\label{FF2}
 \begin{array}{l}
     \displaystyle \mathcal{F}_1(\bx):= \nabla_{\bx}^2\hat{\phi}_1+(\hat{V}(\bx)\hat{\phi}_1-\hat{W}(\bx)\hat{\phi}_2)+\frac{g(\hat{\phi}_1^2+\hat{\phi}_2^2)}{1+S(\hat{\phi}_1^2+\hat{\phi}_2^2)}\hat{\phi}_1-\mu\hat{\phi}_1,\v\\
     \displaystyle \mathcal{F}_2(\bx):= \nabla_{\bx}^2\hat{\phi}_2+(\hat{V}(\bx)\hat{\phi}_2+\hat{W}(\bx)\hat{\phi}_1)+\frac{g(\hat{\phi}_1^2+\hat{\phi}_2^2)}{1+S(\hat{\phi}_1^2+\hat{\phi}_2^2)}\hat{\phi}_2-\mu\hat{\phi}_2,
 \end{array}
\end{equation}
and proceed by approximating $\phi(\bx)$ and $U(\bx)=V(\bx)+iW(\bx)$ by a complex-valued deep neural network. And they can be trained by minimizing the mean squared error loss containing three parts
\begin{equation}\label{lloss}
  \mathcal{T\!L}=\mathcal{L}_{\tilde{f}}+\mathcal{L}_{\phi}+\mathcal{L}_{UB},
\end{equation}
with
\begin{equation}\label{llfib}
 \begin{array}{rl}
     \displaystyle\quad \mathcal{L}_{\tilde{f}}
     =&\d \frac{1}{N_f}\sum_{\ell=1}^{N_f}\left(|\mathcal{F}_1(\bx_f^\ell)|^2
     +|\mathcal{F}_2(\bx_f^\ell)|^2\right)+\frac{1}{N_f}\sum_{\ell=1}^{N_f}\left(|\hat{\phi}_1(\bx_f^\ell)
     -\phi_1^\ell|^2+|\hat{\phi}_2(\bx_f^\ell)-\phi_2^\ell|^2\right)\v\\
     & \d+\frac{1}{N_B}\sum_{\ell=1}^{N_B}\left(|\hat{V}(\bx_B^\ell)-V(\bx_B^\ell)|^2
     +|\hat{W}(\bx_B^\ell)-W(\bx_B^\ell)|^2\right),
 \end{array}
\end{equation}
where $\{\bx_f^\ell,\phi_1^\ell,\phi_2^\ell\}_\ell^{N_f}$ are connected with the training data on the real part and imaginary part of exact solution of Eq.~(\ref{ode}) with $\phi(\bx_f^\ell)=\phi_1^\ell+i\phi_2^\ell$ in domain $\Omega$, and $\{\bx_B^\ell,V(\bx_B^\ell),W(\bx_B^\ell)\}_\ell^{N_B}$ are linked with the randomly selected boundary training data of potential $U(\bx)=V(\bx)+iW(\bx)$ in domain $\partial\Omega$. Finally, when $loss<\varepsilon$ where $\varepsilon$ is an upper bound of loss, the predicted potential $\hat{U}(\bx)=\hat{V}(\bx)+i\hat{W}(\bx)$ is output (see Fig.~\ref{net3} for more details).

The major steps of the PINNs method determining the $\PT$-symmetric potentials of the corresponding stationary equation are
given in Table~\ref{table-pinns2}.
\begin{table}[!t]
\centering
\caption{The PINNs method learning $\PT$-symmetric potential of the stationary equation (\ref{ode}).}
\begin{tabularx}{\textwidth}{lX}
\hline\hline
Step &   Instruction \\
\hline
1 & Constructing a fully-connected neural network NN$(\bx; W, B)$ with initialized parameters $W = \{w_j\}_{1}^{n+1}$ and $B = \{b_j\}_1^{n+1}$ being the weights and bias, and the PINNs $\mathcal{F}(\bx)$ is given by Eq.~(\ref{FF}); \\
2 &  Construct the two training data sets in domain $\Omega\times[0,Z]$ and boundary $\partial\Omega\times[0,Z]$
for the boundary value and considered model;\\
  3&  Constructing a training loss function $\mathcal{T\!L}$ given by Eq.~(\ref{lloss}) by summing the MSE containing three parts $\mathcal{L}_{\tilde{f}}$, $\mathcal{L}_{\phi}$ and $\mathcal{L}_{UB}$;\\
4 &  Training the NN to optimize the parameters $\{W, B\}$ by minimizing the loss function $\mathcal{TL}$ in terms of
the Adam \& L-BFGS optimization algorithm.\\
\hline\hline
\end{tabularx}
\label{table-pinns2}
\end{table}

In the following, we use the above two types of network structures to discover data-driven $\PT$-symmetric potentials of the NLSE (\ref{nls}) with SN in 1D and 2D cases, and compare them from different aspects.

\subsection{Data-driven $\PT$ Scarf-II potential discovery of the SNLSE}

 In this subsection, we will use the above-mentioned PINNs and mPINNs to learn the inverse problems for the $\PT$ non-periodic Scarf-II potentials in 1D and 2D SNLSEs.

\subsubsection{1D SNLSE with $\PT$ Scarf-II potential}

In this subsection, we study the data-driven Scarf-II potential discovery in the 1D SNLSE. Since we are talking about the stationary solution, the two network structures are both used and compared hereinafter.

\begin{figure*}[!t]
    \centering
\vspace{-0.15in}
  {\scalebox{0.7}[0.8]{\includegraphics{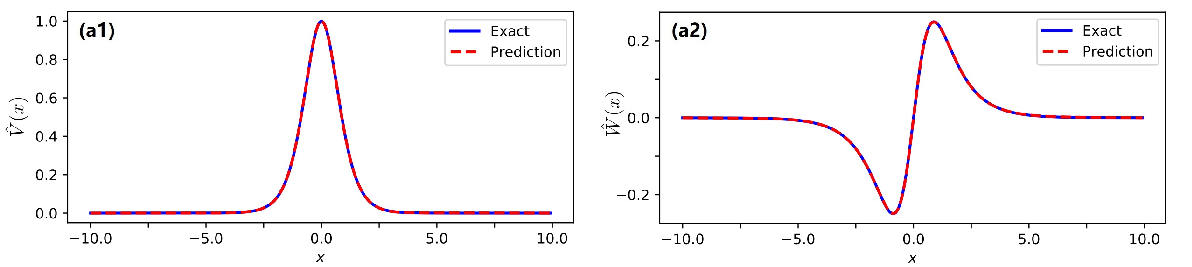}}}\hspace{-0.35in}
\vspace{0.1in}
\caption{\small The data-driven 1D Scarf-II potential of the SNLSE by PINNs related to stationary equation (\ref{ode}). (a1) The comparison of the real part $V(x)$ between the learning and exact Scarf-II potential. (a2) The comparison of the imaginary part $W(x)$ between the learning and exact Scarf-II potential.}
  \label{fig5}
\end{figure*}

\v {\it Case 1.}---For the PINNs related to stationary equation (\ref{ode}) with 1D Scarf-II potential, we firstly generate a training data-set by randomly choosing $N_f=200$ points in the solution region, obtained by the similar way in Sec. \ref{sec2} with $V_0 = 1$, $W_0 = 0.5$, $\mu=1$ and $x\in[-10,10]$, and take $N_B=2$ at the boundary. Then the obtained data-set is applied to train a 3-hidden-layer deep neural network with 32 neurons per layer and a same hyperbolic tangent activation function to approximate the potential $U(x)=V(x)+iW(x)$ in terms of minimizing the mean squared error loss given by Eqs.~(\ref{lloss}) and (\ref{llfib}). Then by using the 10000 steps Adam and 10000 steps L-BFGS optimizations, we obtain the learning potential $\hat{U}(x)$ whose real and imaginary parts respectively are shown in Figs.~\ref{fig5}(a1, a2). The exact potential is also given in Fig.~\ref{fig5}. The relative $\mathbb{L}^2$ norm errors of $V(x)$ and $W(x)$ respectively are $2.647 \cdot 10^{-3}$ and $5.134 \cdot 10^{-3}$. And the learning times of Adam and L-BFGS optimizations are 106s and 113s, respectively.

\begin{figure*}[!t]
    \centering
\vspace{-0.15in}
  {\scalebox{0.7}[0.8]{\includegraphics{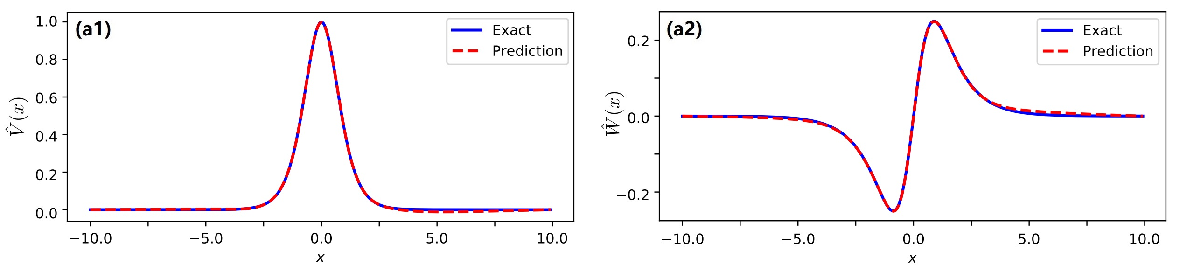}}}\hspace{-0.35in}
\vspace{0.1in}
\caption{\small The data-driven 1D Scarf-II potential of the SNLSE by mPINNs. (a1) The comparison of the real part $V(x)$ between the learning and exact Scarf-II potential. (a2) The comparison of the imaginary part $W(x)$ between the learning and exact Scarf-II potential.}
  \label{fig5p}
\end{figure*}

\v {\it Case 2.}---For the mPINNs related to the SNLSE (\ref{nls}) with 1D Scarf-II potential, we also form a training data-set by randomly choosing $N_f=200$ points in the solution region arising from the numerical solution $\psi(x,z)=\phi(x)e^{i\mu z}$ obtain by Fourier spectral method with $V_0 = 1$, $W_0 = 0.5$ and $(x,z)\in[-10,10]\times[0,1]$, and randomly take $N_B=2$ at the boundary. And we choose the same network i.e., a 3-hidden-layer deep neural network with 32 neurons per layer, and use 10000 steps Adam and 10000 steps L-BFGS optimizations to minimize the mean squared error loss given by Eqs.~(\ref{mloss}) and (\ref{mlfib}). Then the predicted potential is obtain as shown in Figs.~\ref{fig5p}(a1, a2). By comparing with the exact solution, we find that the predicted potential does not fit very well. And the relative $\mathbb{L}^2$ norm errors of $V(x)$ and $W(x)$ respectively are $1.886 \cdot 10^{-2}$ and $3.500 \cdot 10^{-2}$. However when increasing the number of domain points and boundary points $N_f=2000$ and $N_B=50$, the learning potential can achieve the same precision only after 8000 steps Adam and 15000 steps L-BFGS optimizations. The relative $\mathbb{L}^2$ norm errors of $V(x)$ and $W(x)$ respectively are $3.159 \cdot 10^{-3}$ and $3.535 \cdot 10^{-3}$.

Table \ref{table1} illustrates the relative $\mathbb{L}^2$ norm errors of learning $V(x)$ and $W(x)$ and the total learning times at different parameters by using mPINNs and PINNs related to stationary equation network structures. In general, using PINNs, training time is faster and fewer data sets are required since the dimension is one less than mPINNs. However, for non-stationary solution, one can only use mPINNs to discuss the data-driven potential. Importantly mPINNs can also achieve the same accuracy by increasing the data sets.

\begin{table}[!t]
\begin{center}
\caption{Comparison of mPINNs(PDE) and PINNs (stationary equation) for 1D Scarf-II potential, their errors and total learning times at different parameters, where the network is a 3-hidden-layer deep neural network with 32 neurons per layer.}\label{table1}
\begin{tabular}{ c|cccccccc }
\hline
\hline
\rule{0pt}{13pt} \diagbox{Cases}{Items}     & $N_f$ & $N_B$ & Z   & Adam & L-BFGS & time & $\mathbb{L}^2$ error of $V(x)$ &  $\mathbb{L}^2$ error of $W(x)$\\[1ex]
\hline
\rule{0pt}{13pt}  PINNs & 200    & 2     & / &  10000  &  10000      &         219s  &   $2.647 \cdot 10^{-3}$ & $5.134 \cdot 10^{-3}$ \\[1ex]
\rule{0pt}{13pt}   mPINNs(PDE) & 200    & 2     & 1 &   10000   &   10000     &        295s   & $1.886 \cdot 10^{-2}$ & $3.500 \cdot 10^{-2}$ \\[1ex]
\rule{0pt}{13pt}   mPINNs(PDE) & 2000    & 50     & 1 &   8000   &   15000     &        487s   & $3.159 \cdot 10^{-3}$ & $3.535 \cdot 10^{-3}$ \\[1ex]
\hline
\hline
\end{tabular}
\end{center}
\end{table}

\subsubsection{1D SNLSE with $\PT$ Scarf-II potential dependent on propagation distance $z$}

Then we investigate the inverse problem about $\PT$ Scarf-II potential dependent on propagation distance $z$ using the mPINNs method.

Especially, when considering the adiabatic excitations of solutions for 1D SNLSE with $\PT$ Scarf-II potential, the potential parameters are changed adiabatically as functions of the propagation distance $z$, that is, $V_0 \rightarrow V_0(z)$ and $W_0 \rightarrow W_0(z)$~\cite{We15, pt6}, which can be achieved by adding the simultaneous adiabatic switch on $\PT$-symmetric potential. Then the 1D SNLSE with $\PT$ Scarf-II potential becomes the following form
\begin{equation}\label{nlsgz}
  i\psi_z+\psi_{xx}+[V(x,z)+iW(x,z)]\psi++\frac{g|\psi|^2\psi}{1+S|\psi|^2}=0,
\end{equation}
where the adiabatically changed parameters $V_0(z)$ and $W_0(z)$ determining $V(x,z)+iW(x,z)$ are all taken in the same functional form
\begin{equation}\label{condition}
\Theta(z)\!=\!\left\{\!
\begin{array}{ll}
\Theta_1, &             0\leq z < z_1, \\
\Theta_1\!+\!(\Theta_2\!-\!\Theta_1)\!\sin\!\left(\!\dfrac{z\!-\!z_1}{2z_1}\pi\!\right), & z_1\leq z < 2z_1, \\
\Theta_2, & 2z_1\leq z,
\end{array}
\right.
\end{equation}
where $\Theta_1$ and $\Theta_2$ represent the parameters of initial final values in excitation, respectively. In particular we denoted $V_{0i}=\Theta_i$ and $W_{0i}=\Theta_i$.

\begin{figure*}[!t]
    \centering
  {\scalebox{0.7}[0.7]{\includegraphics{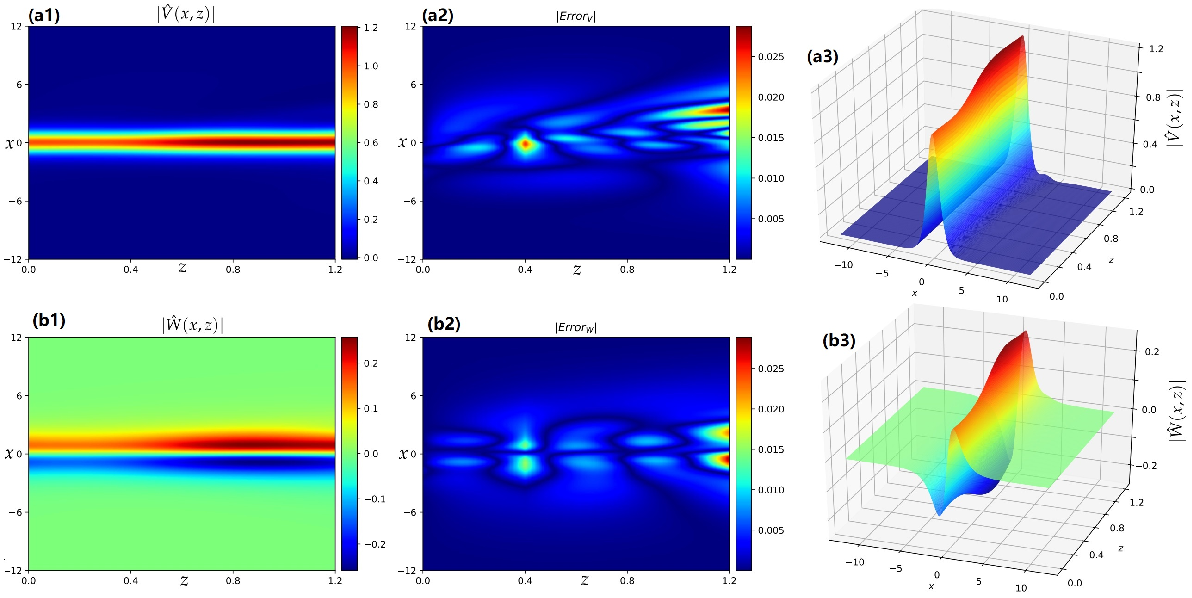}}}\hspace{-0.35in}
\vspace{0.1in}
\caption{\small  The data-driven 1D Scarf-II potential dependent on propagation distance $z$ of Eq. (\ref{nlsgz}) by mPINNs method. (a1, a3) The learning real part $\hat{V}(x,z)$ of potential. (b1, b3) The learning imaginary part $\hat{W}(x,z)$ of potential. (a2) The absolute values of the error between the real part of exact and learning potential $Error_V = |\hat{V}-V|$.
(b2) The absolute values of the error between the imaginary part of exact and learning potential $Error_W = |\hat{W}-W|$.}
  \label{fig5e}
\end{figure*}

For the mPINNs related to Eq.~(\ref{nlsgz}), we delete the $m\mathcal{L}_{UZ}$ from the loss function (\ref{mloss}) because the potential is dependent on propagation distance $z$.
Firstly we form a training data-set by randomly choosing $N_f = 5000$ points in the solution region arising from the evolution with $V_{01}=1, V_{02}=1.2$ and $W_{01} = 0.3, W_{02} = 0.5$  and $(x, z) \in [-12, 12] \times [0, 1.2] (z_1=0.4)$  and randomly take $N_B = 50$ at the boundary.
And we choose a 3-hidden-layer deep neural network with 32 neurons per layer, and use 15000 steps Adam and 30000 steps L-BFGS optimizations to minimize the mean squared error loss. Then the predicted potential is obtain whose real and imaginary parts respectively are shown in Figs.~\ref{fig5e}(a3, b3). The intensity of the real and imaginary parts are also displayed in Figs.~\ref{fig5e}(a1, b1). And the absolute values of the errors between the real and imaginary parts of exact and learning potential $Error_V = |\hat{V}-V|$, $Error_W = |\hat{W}-W|$ are also calculated which are exhibited in Figs.~\ref{fig5e}(a2, b2). The relative $\mathbb{L}^2$ norm errors of $V(x,z)$ and $W(x,z)$ respectively are $1.241 \cdot 10^{-2}$ and $4.267 \cdot 10^{-2}$. And the learning times of Adam and L-BFGS optimizations are 671s and 1241s, respectively.

\subsubsection{2D SNLSE with $\PT$ Scarf-II potential}

In the following, we discuss the data-driven $\PT$ Scarf-II potential in the 2D SNLSE.

\begin{figure*}[!t]
    \centering
  {\scalebox{0.7}[0.7]{\includegraphics{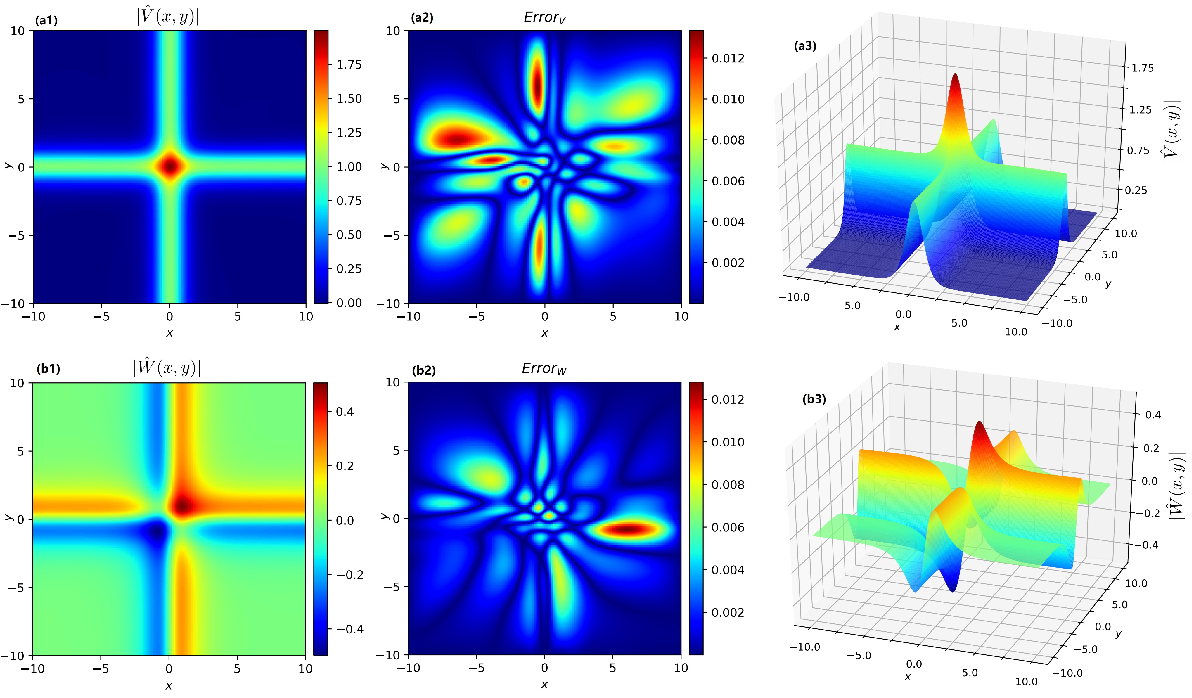}}}\hspace{-0.35in}
\vspace{0.05in}
\caption{\small The data-driven 2D Scarf-II potential of the SNLSE by PINNs related to stationary equation. (a1, a3) The learning real part $\hat{V}(x,y)$ of potential. (b1, b3) The learning imaginary part $\hat{W}(x,y)$ of potential. (a2) The absolute values of the error between the real part of exact and learning potential $Error_V = |\hat{V}-V|$.
(b2) The absolute values of the error between the imaginary part of exact and learning potential $Error_W = |\hat{W}-W|$. }
  \label{fig6}
\end{figure*}

\v {\it Case 1.}---For the PINNs related to stationary equation (\ref{ode}) with 2D Scarf-II potential, we firstly generate a training data-set by randomly choosing $N_f=5000$ points in the solution region arising from the numerical solution $\phi(x,y)$ obtained by numerical method with $W_0 = 0.5$, $\mu=1$ and $x\in[-10,10]\times[-10,10]$ and randomly take $N_B=200$ at the boundary. Then the obtained data-set is applied to train a 3-hidden-layer deep neural network with 32 neurons per layer and a same hyperbolic tangent activation function to approximate the potential $U(x,y)=V(x,y)+iW(x,y)$ in terms of minimizing the mean squared error loss given by Eqs.~(\ref{lloss}) and (\ref{llfib}). Then by using the 10000 steps Adam and 30000 steps L-BFGS optimizations, we obtain the learning potential $\hat{U}(x,y)$, whose real and imaginary parts respectively are shown in Figs.~\ref{fig6}(a3, b3). The intensities of the real and imaginary parts are also displayed in Figs.~\ref{fig6}(a1, b1). And the absolute values of the errors between the real and imaginary parts of exact and learning potential $Error_V = |\hat{V}-V|$, $Error_W = |\hat{W}-W|$ are also calculated which are exhibited in Figs.~\ref{fig6}(a2, b2). The relative $\mathbb{L}^2$ norm errors of $V(x,y)$ and $W(x,y)$ respectively are $9.471 \cdot 10^{-3}$ and $1.868 \cdot 10^{-2}$. And the learning times of Adam and L-BFGS optimizations are 318s and 985s, respectively.

\begin{figure*}[!t]
    \centering
  {\scalebox{0.7}[0.7]{\includegraphics{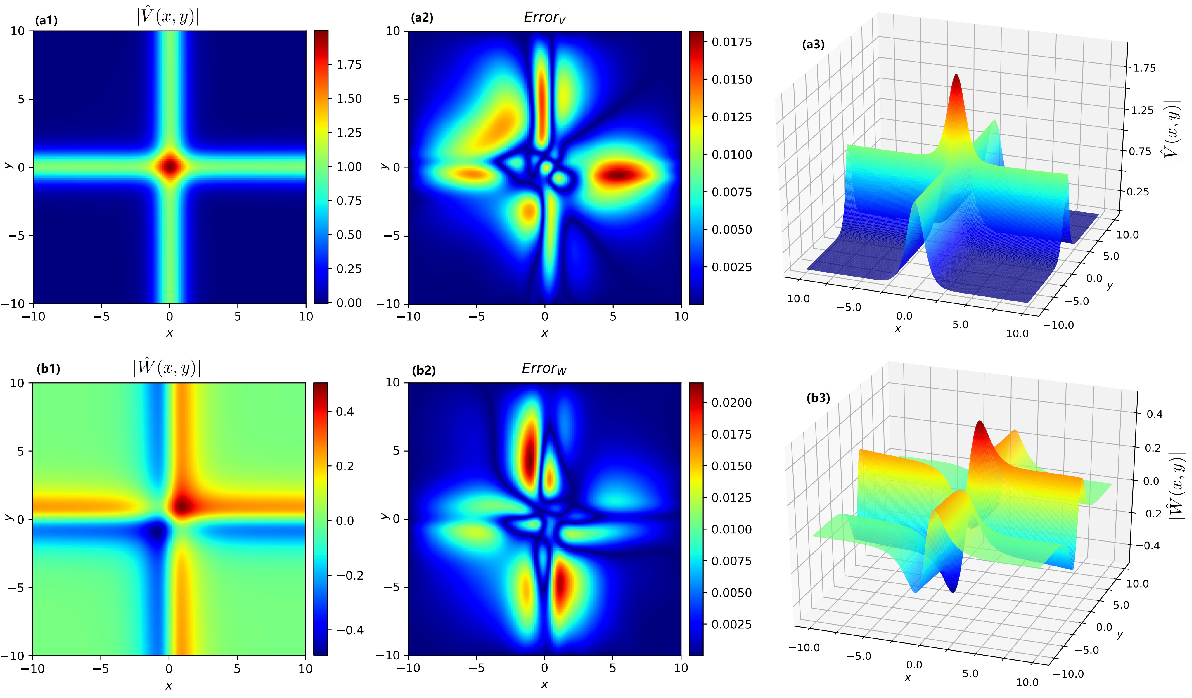}}}\hspace{-0.35in}
\vspace{0.05in}
\caption{\small The data-driven 2D Scarf-II potential of the SNLSE  by mPINNs. (a1, a3) The learning real part $\hat{V}(x,y)$ of potential. (b1, b3) The learning imaginary part $\hat{W}(x,y)$ of potential. (a2) The absolute values of the error between the real part of exact and learning potential $Error_V = |\hat{V}-V|$.
(b2) The absolute values of the error between the imaginary part of exact and learning potential $Error_W = |\hat{W}-W|$. }
  \label{fig6p}
\end{figure*}

\v {\it Case 2.}---For the mPINNs related to SNLSE (\ref{nls}) with 2D Scarf-II potential, we also form a training data-set by randomly choosing $N_f=5000$ points in the solution region arising from the numerical solution $\phi(x,y)$ obtained by numerical method with $W_0 = 0.5$ and $(x,y,z)\in\Omega\times[0,1]$ where $\Omega=[-10,10]\times[-10,10]$ and randomly take $N_B=200$ at the boundary. And we choose the same network i.e., a 3-hidden-layer deep neural network with 32 neurons per layer, and use 10000 steps Adam and 30000 steps L-BFGS optimizations to minimize the mean squared error loss given by Eqs.~(\ref{mloss}) and (\ref{mlfib}). Similar results are also obtained as shown in Figs.~\ref{fig6p}.
By the absolute values of the errors exhibited in Figs.~\ref{fig6p}(a2, b2), we observe that the errors are not much different from that in PINNs related to stationary equation. And the relative $\mathbb{L}^2$ norm errors of $V(x,y)$ and $W(x,y)$ respectively are $1.329 \cdot 10^{-2}$ and $4.083 \cdot 10^{-2}$. Besides, we further increase the number of domain points and boundary points $N_f=15000$ and $N_B=300$. The error of learning potential can be further reduced after 15000 steps Adam and 40000 steps L-BFGS optimizations. Then the relative $\mathbb{L}^2$ norm errors of $V(x,y)$ and $W(x,y)$ respectively are $8.294 \cdot 10^{-3}$ and $1.964 \cdot 10^{-2}$.

\begin{figure}[!h]
    \centering
\vspace{-0.15in}
  {\scalebox{0.7}[0.85]{\includegraphics{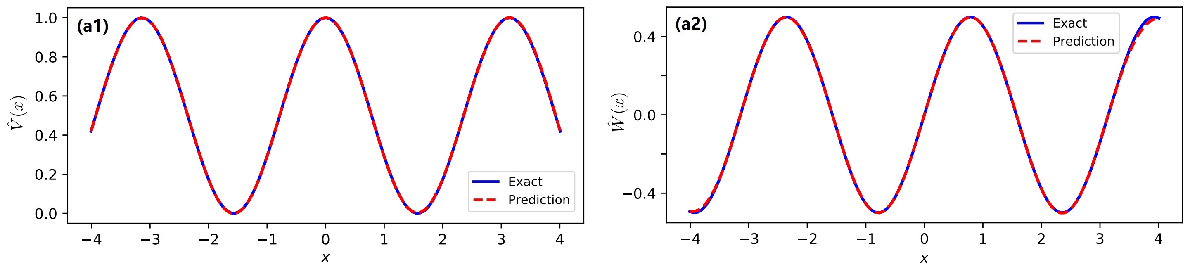}}}
\vspace{0.15in}
\caption{\small The data-driven 1D $\PT$ periodic potential of the SNLSE by mPINNs. (a1) The
comparison of the real part $V(x)$ between the learning and exact $\PT$ periodic potential. (a2) The comparison of the imaginary part
$W(x)$ between the learning and exact $\PT$ periodic potential.}
  \label{fig7}
\end{figure}

Table \ref{table2} exhibits the relative $\mathbb{L}^2$ norm errors of learning $V(x,y)$ and $W(x,y)$ and the total learning times at different parameters by using mPINNs and PINNs related to stationary equation network structures. We observe that they can achieve the same precision for the data-driven Scarf-II potential.

\begin{table}
\begin{center}
\caption{Comparison of mPINNs(PDE) and PINNs (stationary equation) for 2D Scarf-II potential, their errors and total learning times at different parameters, where the network is a 3-hidden-layer deep neural network with 32 neurons per layer.}\label{table2}
\begin{tabular}{ c|cccccccc }
\hline
\hline
\rule{0pt}{13pt}  \diagbox{Cases}{Items}           & $N_f$ & $N_B$ & Z   & Adam & L-BFGS & time & $\mathbb{L}^2$ error of $V(\bx)$ & $\mathbb{L}^2$ error of $W(\bx)$ \\[1ex]
\hline
\rule{0pt}{13pt}  PINNs & 5000    & 200     & / &  10000  &  30000      &         1303s  &   $9.471 \cdot 10^{-3}$ & $1.868 \cdot 10^{-2}$ \\[1ex]
\rule{0pt}{13pt}   mPINNs(PDE) & 5000    & 200    & 1 &   10000   &   30000     &        1558s   & $1.329 \cdot 10^{-2}$ & $4.083\cdot 10^{-2}$ \\[1ex]
\rule{0pt}{13pt}   mPINNs(PDE) & 15000    & 300     & 1 &   15000   &   40000     &        5472s   & $8.294\cdot 10^{-3}$ & $1.964 \cdot 10^{-2}$ \\[1ex]
\hline
\hline
\end{tabular}
\end{center}
\end{table}

\begin{figure*}[!t]
    \centering
  {\scalebox{0.7}[0.85]{\includegraphics{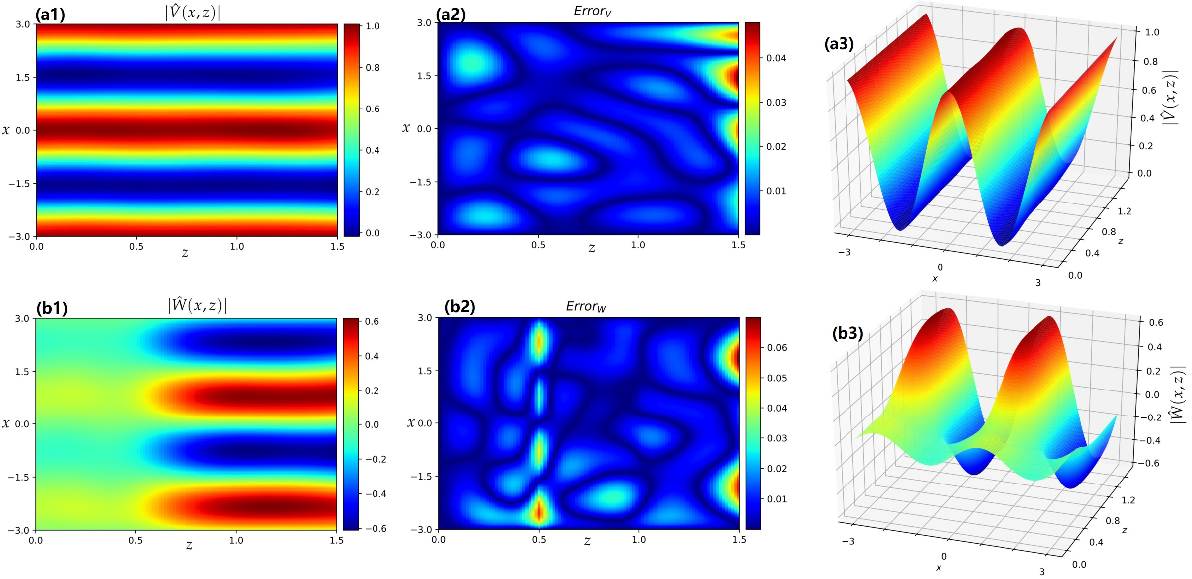}}}\hspace{-0.35in}
\vspace{0.1in}
\caption{\small The data-driven 1D $\PT$ period potential dependent on propagation distance $z$ of Eq. (\ref{nlsgz}) by mPINNs method. (a1, a3) The learning real part $\hat{V}(x,z)$ of potential. (b1, b3) The learning imaginary part $\hat{W}(x,z)$ of potential. (a2) The absolute values of the error between the real part of exact and learning potential $Error_V = |\hat{V}-V|$.
(b2) The absolute values of the error between the imaginary part of exact and learning potential $Error_W = |\hat{W}-W|$.}
  \label{fig7e}
\end{figure*}

\begin{figure*}[!h]
    \centering
\vspace{-0.15in}
  {\scalebox{0.7}[0.7]{\includegraphics{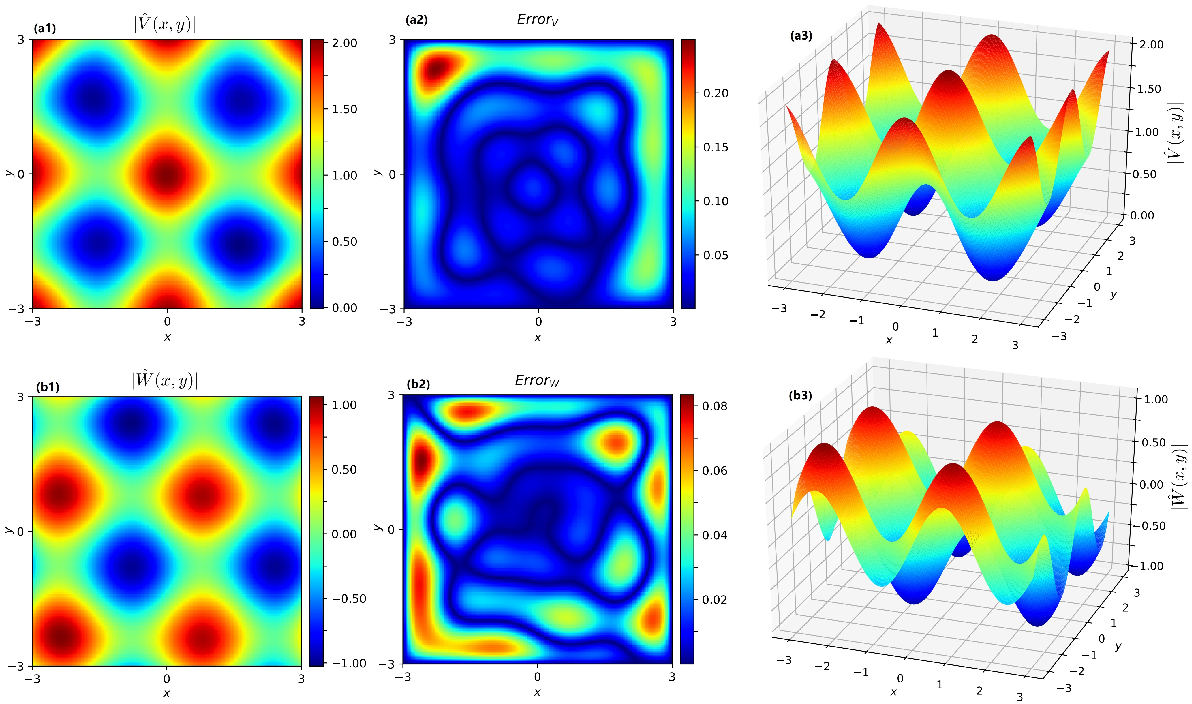}}}\hspace{-0.35in}
\vspace{0.15in}
\caption{\small The data-driven 2D $\PT$ periodic potential of the SNLSE by mPINNs. (a1, a3) The learning real part $\hat{V}(x,y)$ of potential. (b1, b3) The learning imaginary part $\hat{W}(x,y)$ of potential. (a2) The absolute values of the error between the real part of exact and learning potential $Error_V = |\hat{V}-V|$.
(b2) The absolute values of the error between the imaginary part of exact and learning potential $Error_W = |\hat{W}-W|$. }
  \label{fig8}
\end{figure*}

\subsection{Data-driven $\PT$ periodic potential discovery of the SNLSE}

 In this subsection, we will use the above-mentioned PINNs and mPINNs to learn the inverse problems for the $\PT$ periodic potentials in 1D and 2D SNLSEs.

\subsubsection{1D SNLSE with $\PT$ periodic potential}

In this subsection, we investigate the data-driven $\PT$ periodic potential in the 1D SNLSE. Here we consider the non-stationary solution as in Sec. \ref{sec2}. Therefore only mPINNs network structure is available.

Similarly, we use Newton conjugate-gradient method to numerically obtain soliton $\phi(x)$ with zero-boundary conditions. Then by the Fourier spectral method in Matlab \cite{spectral} (i.e., one can take Fourier transform in $x$ space, and choose the explicit fourth-order Runge–Kutta method in propagation distance $z$) to simulate the SNLSE (\ref{nls}) with the initial value condition $\bar{\phi}(x)=\phi(x)e^{-i0.5x}$ and zero-boundary conditions.  Therefore we can generate a .mat data-set about $\psi(x, z)$ in the domain $\Omega \times [0, Z]$ with the 512 Fourier modes in the $x$ direction and propagation-step $\Delta z = 0.005$ as the training data-set. From Sec. \ref{sec2}, we know that it is a non-stationary solution.

For the mPINNs related to PDE (\ref{nls}) with 1D $\PT$ periodic potential, we form a training data-set by randomly choosing $N_f=2000$ points in the solution region arising from the numerical solution evolution by Fourier spectral method with $W_0 = 0.5$ and $(x,z)\in[-4,4]\times[0,1]$ and randomly take $N_B=50$ at the boundary. And we choose  a 3-hidden-layer deep neural network with 32 neurons per layer, and use 15000 steps Adam and 25000 steps L-BFGS optimizations to minimize the mean squared error loss given by Eqs.~(\ref{mloss}) and (\ref{mlfib}). {\color{red}Then the predicted potential is obtained} as shown in Figs.~\ref{fig7}(a1, a2). And the relative $\mathbb{L}^2$ norm errors of $V(x)$ and $W(x)$ respectively are $3.240 \cdot 10^{-3}$ and $1.392 \cdot 10^{-2}$. And the learning times of Adam and L-BFGS optimizations are 331s and 581s, respectively.

\subsubsection{2D SNLSE with $\PT$ period potential dependent on propagation distance $z$}

Then, we consider the inverse problem about $\PT$ period potential dependent on propagation distance $z$ by mPINNs method. Analogously, we change the potential parameter $W_0$ as a function of the propagation distance $z$, that is, $W_0 \rightarrow W_0(z)$ taken in the functional form of Eq. (\ref{condition}).

For the mPINNs related to Eq. (\ref{nlsgz}), we firstly form a training data-set by randomly choosing $N_f = 4000$ points in the solution region arising from the evolution with $W_{01} = 0.1, W_{02} = 0.6$  and $(x, z) \in [-3, 3] \times [0, 1.5] (z_1=0.5)$  and randomly take $N_B = 50$ at the boundary.
And we choose the same network i.e., a 3-hidden-layer deep neural network with 32 neurons per layer, and use 10000 steps Adam and 30000 steps L-BFGS optimizations to minimize the mean squared error loss. Then the predicted potential is obtained, whose real and imaginary parts respectively are shown in Figs.~\ref{fig7e}(a3, b3). The intensities of real and imaginary parts are also displayed in Figs.~\ref{fig7e}(a1, b1). And the absolute values of the errors between the real and imaginary parts of exact and learning potential $Error_V = |\hat{V}-V|$, $Error_W = |\hat{W}-W|$ are also calculated which are exhibited in Figs.~\ref{fig7e}(a2, b2). The relative $\mathbb{L}^2$ norm errors of $V(x,z)$ and $W(x,z)$ respectively are $1.143 \cdot 10^{-2}$ and $4.181 \cdot 10^{-2}$. And the learning times of Adam and L-BFGS optimizations are 284s and 887s, respectively.

\subsubsection{2D SNLSE with $\PT$ periodic potential}

Finally, we discuss the data-driven $\PT$ periodic potential in the 2D SNLSE.
For the mPINNs related to  SNLSE (\ref{nls}) with 2D $\PT$ periodic potential, we firstly generate a training data-set by randomly choosing $N_f=15000$ points in the solution region arising from numerical solution $\psi(x,y,z)$ by Fourier spectral method with $W_0 = 0.5$ and $x\in[-3,3]\times[-3,3]$ and randomly take $N_B=300$ at the boundary. Then the obtained data-set is applied to train a 3-hidden-layer deep neural network with 32 neurons per layer and a sine activation function to approximate the potential $U(x,y)=V(x,y)+iW(x,y)$ in terms of minimizing the mean squared error loss given by Eqs.~(\ref{mloss}) and (\ref{mlfib}). Then by using the 15000 steps Adam and 40000 steps L-BFGS optimizations, we obtain the learning potential $\hat{U}(x,y)$, whose real and imaginary parts respectively are shown in Figs.~\ref{fig8}(a3, b3). The intensities of the real and imaginary parts are also displayed in Figs.~\ref{fig8}(a1, b1). And the absolute values of the errors between the real and imaginary parts of exact and learning potential $Error_V = |\hat{V}-V|$, $Error_W = |\hat{W}-W|$ are also calculated which are exhibited in Figs.~\ref{fig8}(a2, b2). The relative $\mathbb{L}^2$ norm errors of $V(x,y)$ and $W(x,y)$ respectively are $6.018 \cdot 10^{-2}$ and $5.799 \cdot 10^{-2}$. And the learning times of Adam and L-BFGS optimizations are 1479s and 3993s, respectively.

\section{Conclusions and discussions}\label{sec5}

In conclusion, we have investigated the one- and two-dimensional optical solitons and the general non-stationary solutions of the nonlinear Schr\"{o}dinger equation (NLSE) with saturable nonlinearity and two kinds of $\PT$-symmetric Scarf-II and periodic potentials in optical fibres via deep learning PINNs. Of greater significance, the inverse problems of $\PT$-symmetric potential discovery rather than just the potential parameters have been discussed. Based on PINNs we proposed one network structure (mPINNs) to identify the potential of NLSE. And the inverse problems about 1D and 2D $\PT$-symmetric potentials depending on propagation distance $z$ are investigated using mPINNs method.
Particularly, for the stationary solution of SNLSE, we can identify the potential of SNLSE directly by PINNs related to the related stationary equation. Furthermore, the two network structures are compared under different parameter conditions. And we observe that the predicted potential can achieve the same accuracy with two different network structures. Furthermore, some main factors affecting the neural network performance are discussed including nonlinear  activation functions, structures of the neural networks and the sizes of the training data. Twelve special classes of activation functions are illustrated in the deep learning the 1D and 2D SNLSEs with the $\PT$ potential. And some examples are shown to illustrate that selecting the activation function according to the form of solution and equation usually can achieve better effect.

On the other hand, for the mPINNs, we will continue to improve and optimize the network structure to improve training speed in future. For example, we may split the mPINNs into two subnetworks, one taking $\bx$ and $z$ as the inputs for training solution of equation, and another only taking $\bx$ as the input for training potential. In this way, we may simplify the loss function. Besides, consider too many optimization goals, the self-adaptation weights can be applied to focus on training the hard parts.





\v \v \noindent {\bf Acknowledgement}

The work  was supported by the National Natural Science Foundation of China  under Grant No. 11925108.






\end{document}